\documentclass{aa} 
\usepackage{natbib}
\bibpunct{(}{)}{;}{a}{}{,} % to follow the A&A style
\usepackage{graphicx}
\usepackage{amsmath}
\usepackage{geometry}
\usepackage{txfonts}
\usepackage{appendix}
\usepackage[utf8]{inputenc}
\usepackage{lscape}
\newcommand{\spit}{{\it Spitzer}} % Spitzer in Italics
\newcommand{\her}{{\it Herschel}} % Herschel in italics
\newcommand{\sofia}{{\it SOFIA}} % SOFIA in italics
\newcommand{\chandra}{{\it Chandra}} % Chandra in italics

\newcommand{\go}{G$_{\rm{UV}}$}
\newcommand{\gx}{G$_{\rm X}$}
\newcommand{\gstar}{G$_{stars}$}
\newcommand{\Lfir}{L$_\text{FIR}$}                                  %L_fir
\newcommand{\fir}{\text{L}_\text{FIR}}                               % L_fir in math mode
\newcommand{\Ltir}{L$_\text{TIR}$}                                  %L_tir
\newcommand{\Lpdr}{L$_{\rm FIR}^{\rm PDR}$}            %L_fir, pdr
\newcommand{\Lpah}{L$_{\rm PAH}$}            %L_pah
\newcommand{\mic}{\,$\mu$m}                                          %micron outside math mode
\newcommand{\cm}{cm$^{-3}$}                                        %cubic centimeters
\newcommand{\escmsr}{${\rm erg~s}^{-1}{\rm cm}^{-2}{\rm sr}^{-1}$} % erg/s/cm2/sr
\newcommand{ \Xco}{{\rm X}$_{\rm CO}$}   %Xco
\newcommand{\arsec}{$^{\prime\prime}$}       % "
\newcommand{\Av}{$A_{\rm{V}}$}    % Av
\newcommand{\Avmax}{$A_{\rm{V}}^{\rm max}$}    % Av,max
\newcommand{\Avdust}{$A_{\rm{V}}^{\rm dust}$}    % Av,dust
\newcommand{\PhiA}{$\Phi_{\rm A}$} %Phi A

\newcommand{\HH}{H$_2$}          %  H2
\newcommand{\HII}{H$\,${\sc ii}}     %  HII
\newcommand{\Halpha}{H$_\alpha$}  %  Halpha
\newcommand{\OIII}{[O$\,${\sc iii}]}    %  OIII
\newcommand{\SIII}{[S$\,${\sc iii}]}    %  SIII
\newcommand{\SIV}{[S$\,${\sc iv}]}    %  SIV
\newcommand{\OI}{[O$\,${\sc i}]} %OI
\newcommand{\CII}{[C$\,${\sc ii}]} %CII
\newcommand{\CI}{[C$\,${\sc i}]} %CI
\newcommand{\NII}{[N$\,${\sc ii}]} % NII
\newcommand{\ArII}{[Ar$\,${\sc ii}]} % ArII
\newcommand{\NeII}{[Ne$\,${\sc ii}]} % NeII
\newcommand{\SiII}{[Si$\,${\sc ii}]} % SiII
\begin{document}

   \title{A milestone toward understanding PDR properties in the extreme environment of LMC-30Dor}

   \subtitle{}

   \author{ M.~Chevance\inst{1,2}
    \and S.~C.~Madden\inst{1}
    \and V.~Lebouteiller \inst{1}
    \and B.~Godard \inst{3}
    \and D.~Cormier \inst{4}
    \and F.~Galliano\inst{1}
    \and S.~Hony \inst{4}
    \and R.~Indebetouw\inst{5,6}
    \and J.~Le Bourlot\inst{3}
    \and M.-Y.~Lee  \inst{1}
    \and F.~Le Petit\inst{3}
    \and E.~Pellegrini\inst{4}
    \and E.~Roueff\inst{3}
    \and R.~Wu\inst{7}
   % \and et al. 
   }

   \institute{Laboratoire AIM, CEA/DSM - CNRS - Universit\'e Paris Diderot, IRFU/Service d'Astrophysique, CEA Saclay, 91191 Gif-sur- Yvette, France. \email{melanie.chevance@cea.fr}
   \and Universit\'e Paris Diderot, Sorbonne Paris Cit\'e, F-75205 Paris, France
    \and LERMA, Observatoire de Paris \& CNRS, 5 place Jules Janssen, 92190 Meudon, France
    \and Institut f\"ur theoretische Astrophysik, Zentrum f\"ur Astronomie der Universit\"at Heidelberg, Albert-Ueberle Str. 2, D-69120 Heidelberg, Germany
    \and Department of Astronomy, University of Virginia, PO Box 3818, Charlottesville, VA 22903, USA
    \and National Radio Astronomical Observatory, Charlottesville, VA 22904, USA 
    \and Department of Astronomy, Graduate School of Science, The University of Tokyo, Bunkyo-ku, Tokyo 113-0033, Japan
    }

   \date{Received ; accepted }

  \abstract
   {More complete knowledge of galaxy evolution requires understanding the process of star formation and interaction between the interstellar radiation field and the interstellar medium (ISM) in galactic environments traversing a wide range of physical parameter space. 
   Here we focus on the impact of massive star formation on the surrounding low metallicity ISM in 30 Doradus in the Large Magellanic Cloud (LMC). 
   A low metal abundance, as is the case of some galaxies of the early universe, results in %an overall lower galactic dust reservoir, hence, 
   less ultra-violet (UV) shielding for the formation of the molecular gas necessary for star formation to proceed.
       The half-solar metallicity gas in this region is strongly irradiated by the super star cluster R136, making it an ideal laboratory to study the structure of the ISM in an extreme environment.
   }
   {Our goal is to construct a comprehensive, self-consistent picture of the density, radiation field, and ISM structure in the most active star-forming region in the LMC, 30 Doradus.
   Our \textit{spatially} resolved study investigates the gas heating and cooling mechanisms, particularly in the photo-dissociation regions where the chemistry and thermal balance are regulated by far-ultraviolet photons (6 eV< h$\nu$ <13.6 eV).
     }
   {We present \her\ observations of far-infrared fine-structure lines obtained with PACS and SPIRE/FTS.
We have combined atomic fine-structure lines from \her\ and \spit\ observations with ground-based CO data to provide diagnostics on the properties and the structure of the gas by modeling it with the Meudon PDR code.
For each tracer we estimate the possible contamination from the ionized gas in order to isolate the PDR component alone.
We derive the spatial distribution of the radiation field, the pressure, the size, and the filling factor of the photodissociated gas and molecular clouds.
   }
   { We find a range of pressure of $\sim 10^5 - 1.7\times10^6$ cm$^{-3}$ K and a range of incident radiation field \go $\sim 10^2 - 2.5\times10^4$ through PDR modeling.
   Assuming a plane-parallel geometry and a uniform medium, we find a total extinction \Avmax\ of $1-3$ mag , which correspond to a PDR cloud size of 0.2 to 3pc, with small CO depth scale of 0.06 to 0.5pc.
   At least 90\% of the \CII\ originates in PDRs in this region, while a significant fraction of the \Lfir\ (up to 70\% in some places) can be associated with an ionized gas component. % that needs to be corrected for the PDR modeling. 
   The high \OIII/\CII\ ratio (2 to 60) throughout the observed map, correlated with the filling factor, reveals the porosity of the ISM in this region, traversed by hard UV photons, surrounding small PDR clumps. 
   We also determine the three dimensional structure of the gas, showing that the clouds are distributed 20 to 80 pc away from the main ionizing cluster, R136.
}
   {}


   \maketitle
%
%________________________________________________________________

\section{Introduction}

Galaxy evolution is dictated by progressive chemical enrichment which is mostly achieved through a succession of star formation episodes.
The effect of metal enrichment on what we observe to be the star formation and interstellar medium (ISM) properties remains elusive despite circumstantial evidence. 
For example, reduced metallicity is expected to have important consequences on the chemistry and the subsequent heating and cooling mechanisms of the gas and dust, directly affecting the transition of the atomic to molecular phase.
In low metallicity environments, the transition between C$^+$/C/CO can be shifted further into the cloud in physical scale, leaving a relatively larger photodissociation region (PDR) and a smaller CO core, compared to more metal-rich environments, such as the Milky Way \citep{Kaufman1999}. 
This effect on the molecular cloud structure manifests itself in an observed low CO luminosity in dwarf galaxies (e.g. \citealt{Cormier2014, Schruba2012}) and requires a higher CO-to-H$_2$ conversion factor, the \Xco\ factor \citep{Schruba2012, Bolatto2013}.
It could also possibly be explained by a higher star formation efficiency.
 
Dwarf galaxies in our local universe are the closest environments we can explore in detail to witness the interplay between star formation and  ISM under low metallicity conditions. 
Large surveys probing the cooling of dwarf galaxies have been possible for the first time with the \her\ Space Observatory (\citealt{Pilbratt2010} ; e.g. the Dwarf Galaxy Survey, DGS ; \citealt{Madden2013}). 
Recent studies taking advantage of the \her\ sensitivity, have modeled the dust and gas properties of a wide range of low metallicity galaxies on integrated galaxy scales (e.g. \citealt{Remy-Ruyer2013, Remy-Ruyer2014, Remy-Ruyer2015, Cormier2012, Cormier2015, Cigan2015}) and find prominent differences between metal-rich and metal poor galaxies. 
For example, from far-infrared (FIR) line ratios, \cite{Cormier2015} have determined that radiation fields over global galaxy scales are harder in star-forming dwarf galaxies, compared to more metal-rich galaxies. 
Furthermore, the filling factor of the ionized gas appears larger relatively to the neutral gas. 
As a consequence of the low metallicity and low extinction in dwarf galaxies, it is possible that a significant fraction of the molecular gas is not traced by CO, but may be residing in the C$^+$ or C$^0$- emitting region for example (referred to as the "dark gas" in \citealt{Wolfire2010}), quantified in low metallicity environments using C$^+$  by \cite{Poglitsch1995} and \cite{Madden1997} and more recently in our Galaxy  by \cite{Langer2014} and \cite{Pineda2014}.

The Large Magellanic Cloud is our closest low metallicity galaxy neighbor (1/2 $Z_\odot$, \citealt{Rolleston2002, Pagel2003}; 50 kpc, \citealt{Walker2012}) allowing us to zoom into the ISM at the spatial resolution of $\sim 12''$ ($\sim 3$ pc) with \her.  
We focus on 30 Doradus (hereafter "30Dor"), which is  the most prominent star-forming region in the LMC and provides the best laboratory to study the impact of a super star cluster (SSC) on the ISM. 
The primary ultraviolet radiation source illuminating this region is the SSC R136, containing 39 O3 stars \citep{Hunter1999}, often considered to be the most extreme star-forming region in the Local Group.
The lower dust abundance of the LMC 
allows for deep penetration of the ionizing radiation, creating extended PDR regions and a more porous environment channeling the UV photons.  
ALMA observations from \cite{Indebetouw2013} have revealed the clumpy structure of the molecular gas in 30Dor, showing small $^{12}$CO filaments and clumps ($\lesssim 1$ pc) covering about 15\% of their map.
We examine the PDR conditions in the neutral atomic gas using mainly \CII\ and \OI\ observed by \her\ in this region to unveil the spatial distribution of the radiation field and the structure of the photodissociated gas and molecular clouds. The fraction of CO-dark gas based on this detailed study will be quantified in a following study (Chevance et al. in prep, hereafter paper II).

Observations of the ionized gas were conducted with \spit\ and studied by \cite{Indebetouw2009}.
They showed in particular that photoionization dominates the ionization structure of the gas over shocks in the \HII\ region around R136.
 A study of the fine structure lines of C$^+$ and O$^0$ in 30Dor has been previously carried out by \cite{Poglitsch1995}, with the Kuiper Airborne Observatory (KAO), at a resolution of $\sim$55\arcsec\ ($\sim 13$ pc). 
 They found that a highly fragmented structure with high density clouds (n = $10^3-10^4$ cm$^{-3}$), of low relative beam filling-factor of CO compared to the PDR (4\% of the clumps volume), bathed in ionized gas could explain the observed ratio \CII/CO (ten times higher than the Galactic value).
Moreover, most of the molecular gas may be present in the PDR, and faint in CO.
Now, the PACS observations provide better spatial resolution than the KAO data and include other important tracers, with an improved signal-to-noise ratio.
 \cite{Pineda2012} have investigated the CO and \CI\ emission observed with the NANTEN2 4-m telescope in a 26\arcsec\ beam, combined with the KAO observations of \CII\  in 30Dor and found likewise a very clumpy medium.

 We present spectroscopic data of 30Dor in Section~\ref{Obs}.
 Section~\ref{Data} describes the observed maps and some preliminary results.
 In Section~\ref{PDR} we use PDR models to determine the physical parameters of the gas in PDR and we study the impact of metallicity and geometry on these parameters.
 We discuss our results and build a comprehensive 3D picture of the region in Section~\ref{discussion}.
 Key results and conclusions are summarized in Section~\ref{Conclusion}.

%__________________________________________________________________

\section{Observations and data preparation}
  \label{Obs}

  \subsection{\her\ PACS spectroscopy}

\begin{figure*}
     \vspace{1cm}

      \centering
      \includegraphics[trim = 19mm 2mm 3.3mm 0mm, clip, width=6cm]{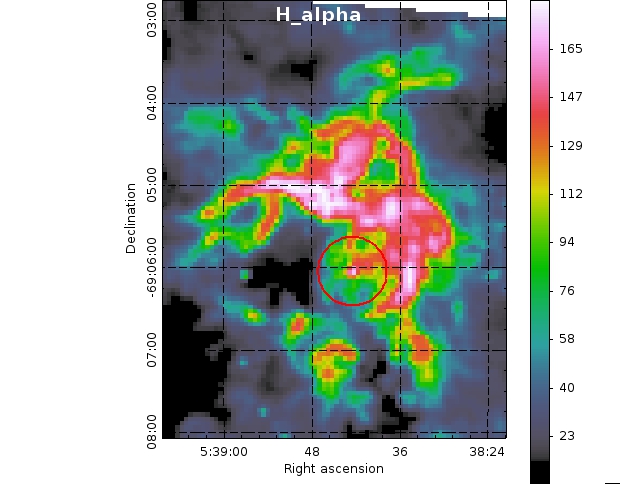} 
      \includegraphics[trim = 19mm 2mm 3.3mm 0mm, clip, width=6cm]{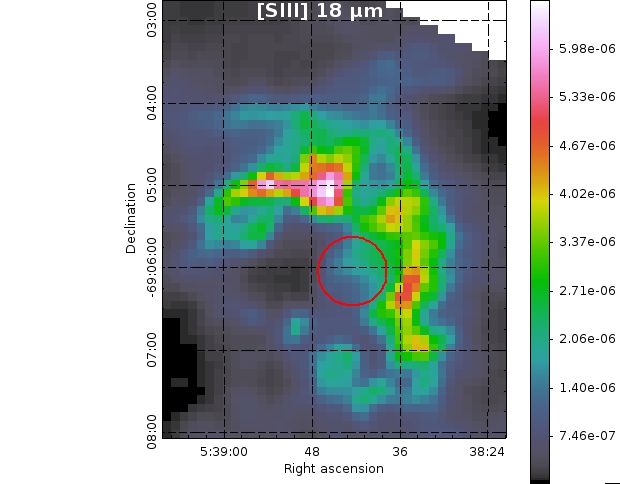}
      \includegraphics[trim = 19mm 2mm 3.3mm 0mm, clip, width=6cm]{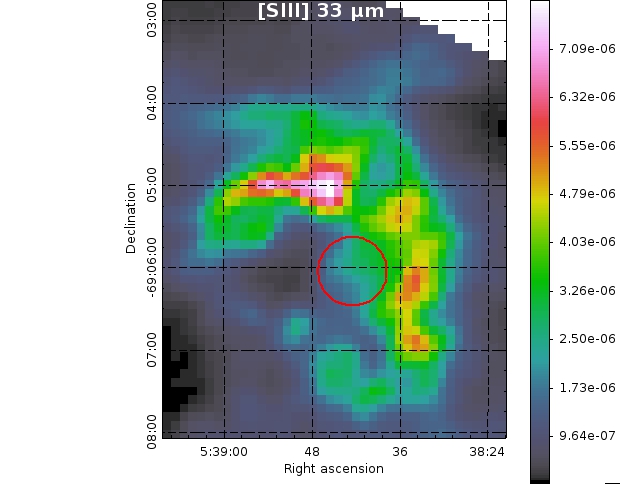}
        
      \includegraphics[trim = 19mm 2mm 3.3mm 0mm, clip, width=6cm]{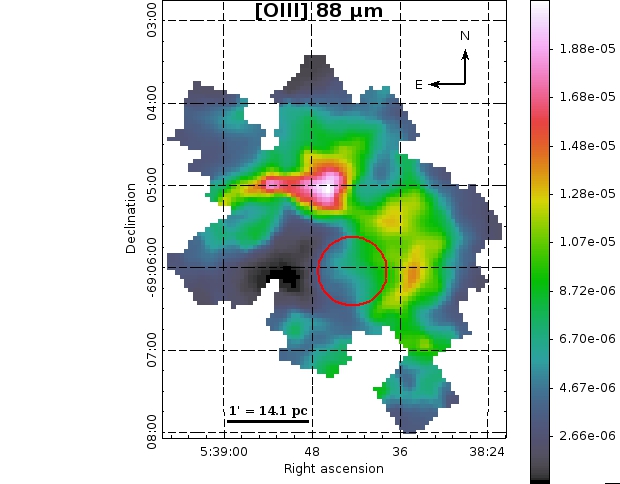}
      \includegraphics[trim = 19mm 2mm 3.3mm 0mm, clip, width=6cm]{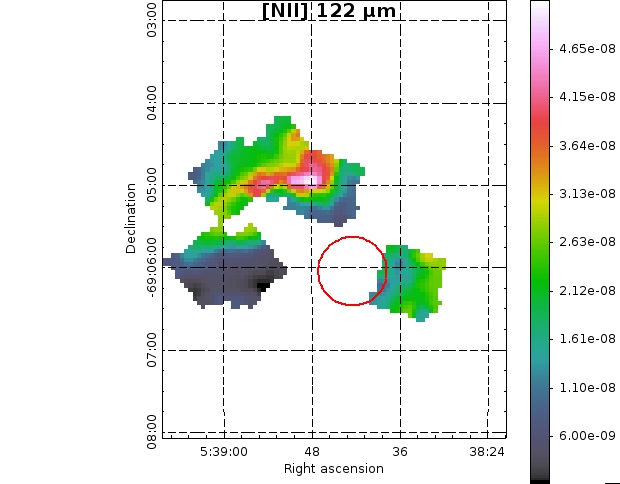}
      \includegraphics[trim = 19mm 2mm 3.3mm 0mm, clip, width=6cm]{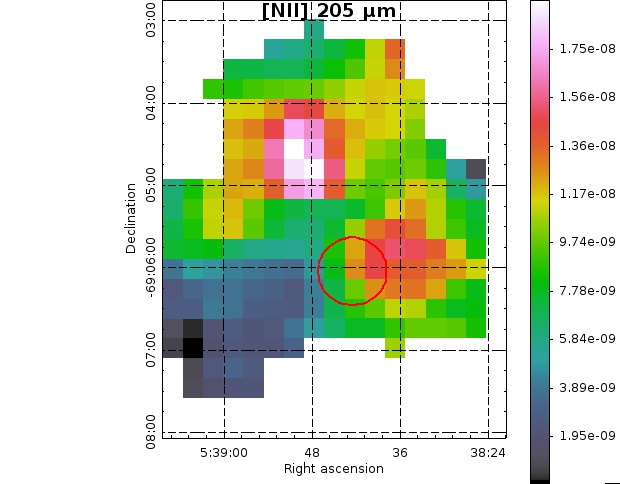}
        
      \includegraphics[trim = 19mm 2mm 3.3mm 0mm, clip, width=6cm]{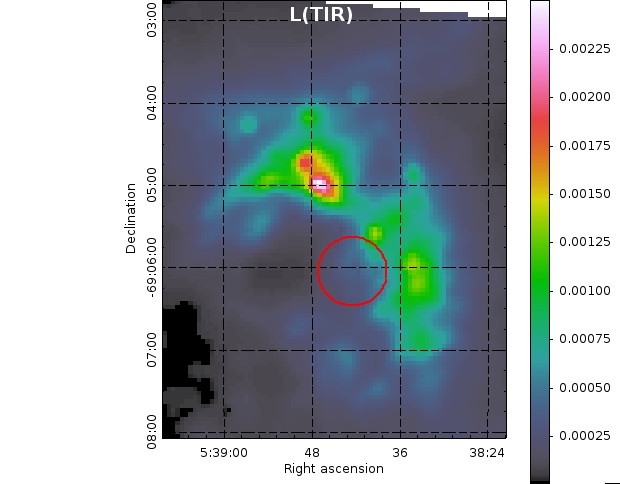}
      \includegraphics[trim = 19mm 2mm 3.3mm 0mm, clip, width=6cm]{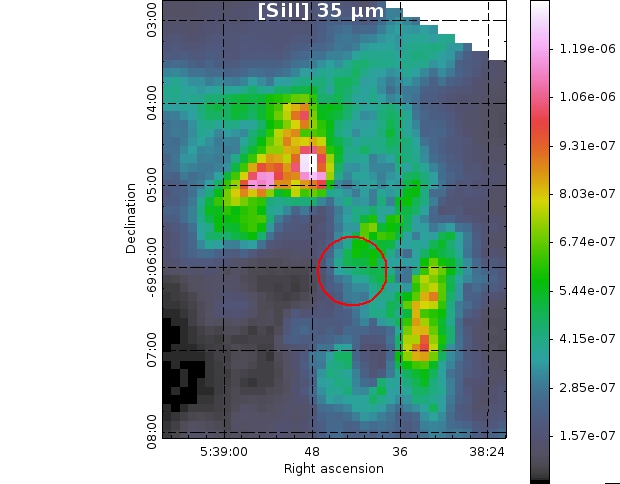}
      \includegraphics[trim = 19mm 2mm 3.3mm 0mm, clip, width=6cm]{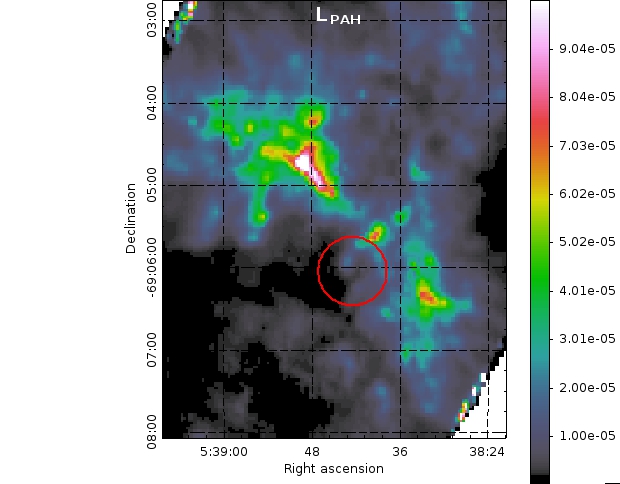} 
      
      \includegraphics[trim = 19mm 2mm 3.3mm 0mm, clip, width=6cm]{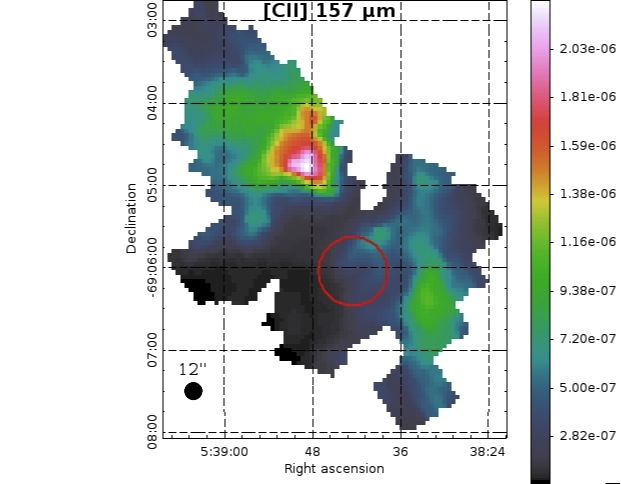}  
      \includegraphics[trim = 19mm 2mm 3.3mm 0mm, clip, width=6cm]{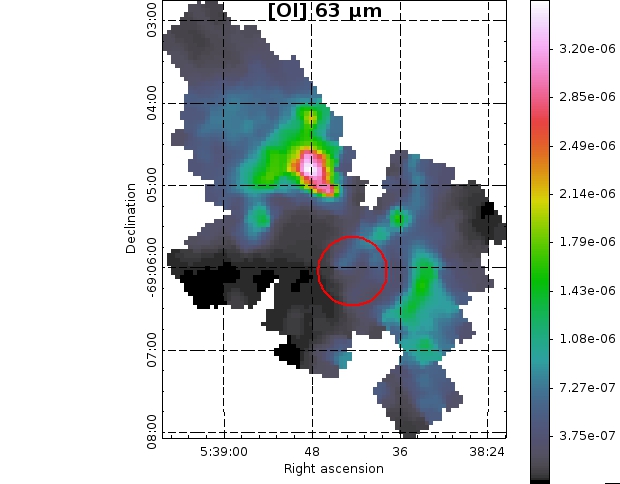}
      \includegraphics[trim = 19mm 2mm 3.3mm 0mm, clip, width=6cm]{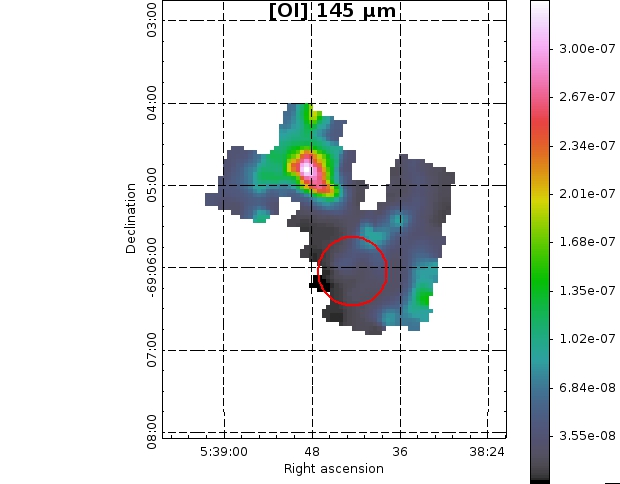}
      
      \caption{PDR and ionized gas line emission from \her\ (PACS and SPIRE/FTS) and \spit /IRS observations of the 30Dor region in W~m$^{-2}$~sr$^{-1}$.The maps are shown at their original spatial resolution. 
      The \Ltir\ map is the total far-infrared luminosity integrated between 3-1000\mic\ from our SED modeling. 
      The red circle represents the location of the R136 cluster.
      Table~\ref{tab:properties} and Section~\ref{Obs} describe these observations in detail.
      } 
      \label{fig:maps}
      \vspace{1cm}
  \end{figure*}

We have mapped five fine structure lines, \CII\ 158\mic , \NII\ 122\mic , \OI\ 63\mic , \OI\ 145\mic\  and \OIII\ 88\mic\  using the Photodetector Array Camera and Spectrometer (PACS, \citealt{Poglitsch2010}) towards 30Dor. 
Properties of these lines are presented in Table~\ref{tab:properties} and the maps can be seen in Figure~\ref{fig:maps}.
These observations, described in \cite{Madden2013}, are part of the \her\ key program, SHINING (P.I. E. Sturm). %, described in \cite{Madden2013}.
We also used two additional pointings east of R136, which were observed by Indebetouw et al (OT2) in \CII, \OI\ 63\mic , \NII\ 122\mic\ and \OIII\ 88\mic .
The details of the observations are shown in Appendix \ref{app:pacs}.

The PACS array is composed of $5 \times 5$ spatial pixels (or \textit{spaxels}) of 9.4$^{\prime \prime}$ covering a total field of view of $47''$. 
The fine structure lines \OI\ 63\mic , \OIII\ 88\mic , \NII\ 122\mic , \OI\ 145\mic\ and \CII\ were mapped with respectively 25, 25, 4, 11 and 31 raster positions, covering approximately a $4^{\prime} \times 5^{\prime}$ region ($56\text{ pc} \times 70\text{ pc}$).
The observations were done in unchopped mode. 
The beam size is 9.5'' at 60 \mic, and 12'' at 160 \mic\ (PACS Observer's Manual 2011).
  
 We refer to \cite{Cormier2015} for the full description of the PACS observations and data reduction, and we summarize here some of the main steps.
  The data were reduced with the \her\ Interactive Processing Environment (HIPE) v12.0.0 \citep{Ott2010} from Level 0 to Level 1. 
  The Level 1 cubes, calibrated in flux and wavelength, are then exported and processed with \mbox{PACSman} v3.61 \citep{Lebouteiller2012} to fit the lines and create the individual maps. 
  Each spectrum is fitted with a second order polynomial for the baseline and a Gaussian for the line. 
  Finally the individual rasters are projected onto a common grid of $3''\times 3''$ pixels  ( $0.72 \times 0.72$ pc) to reconstruct the final maps.
  Uncertainties on the fit and on the projection are estimated using a Monte-Carlo simulation. 
  All of the lines are well detected everywhere in the map (Fig.~\ref{fig:maps}).
  The weakest line, \NII\ 122\mic\ has a signal-to-noise ratio (SNR) between 5 and 30 for most of the mapped area.
  The emission line \OI\ 145\mic\ has a SNR between 7 and 90 and the SNR is above 10 for all of the other lines.

 The observed intensities match well those detected with the Kuiper Airborne Observatory by \cite{Poglitsch1995} with a lower spatial resolution (55\arsec\ for \CII\ and \OI\ 145\mic\ and 22\arsec\ for \OI\ 63\mic ).
 For example they found a maximum \CII\ intensity of $10^{-3}$ \escmsr , and we measure a maximum intensity of $1.1\times10^{-3}$ \escmsr\ on PACS data convolved to the same resolution.
 They are also similar to the \CII\ intensities measured by Requena-Torres (in prep) using the GREAT instrument on \sofia .

     \begin{table*}
       \centering
       \caption{Properties of lines and observations.}
       \begin{tabular}{l l c c c c}
	  \hline
	  \hline
	  Instrument & Transition                            & $\lambda$ (\mic) & FWHM (arcsec) & Ionization energy (eV)      & $n_{crit}$\tablefootmark{a} (cm$^{-3}$)  \\ 
	  \hline
	  
	 PACS          & \OI\ $^3P_{1} -  ^3P_{2}$        &63.2                        & 9.5                       &  --                            & $4.7 \times 10^5$ [H]      \\
	  
	                      &\OI\ $^3P_{0} - ^3P_{1}$        & 145.5                      & 11.0                     &  --                            & $9.5 \times 10^4$ [H]      \\
	  
	                      &\OIII\ $^3P_{1} - ^3P_{0}$         & 88.3                       & 9.5                        &  35.1                       & $510$ [e]     \\
	 
	                      &\CII\ $^2P_{3/2} - ^2P_{1/2}$  & 157.7                     &  11.6                     &11.3                       & $2.8 \times 10^3$ [H], 50 [e]       \\
	  
	                      &\NII\ $^3P_{2} - ^3P_{1}$         & 121.8                     & 9.9                        & 14.5                       & $310$ [e]       \\
	  \hline
	  SPIRE/FTS&\NII\ $^3P_{1} - ^3P_{0}$        & 205.3                    & 16.6                      &  14.5                      & $48 $ [e]     \\
	  
	                      &\CI\ $^3P_{2} - ^3P_{1}$         &  370.4                       & 36.2                       &  --                           & $1.2 \times 10^3$ [\HH]     \\
	   
	                      &\CI\ $^3P_{1} - ^3P_{0}$         & 609.7                        & 38.6                       &  --                           & $4.7 \times 10^3 $ [\HH]     \\
	  \hline
	 Spitzer/IRS\tablefootmark{b}&\SIII\ $^3P_{2} - ^3P_{1}$ & 18.7  & 4.9                         &  23.3                      & $2 \times 10^4 $ [e]      \\  %with electrons
	  
	                     &\SIII\ $^3P_{1} - ^3P_{0}$          & 33.5                      & 8.9                         &  23.3                      & $7 \times 10^3 $ [e]     \\
	                     
	                     &\SiII\  $^2P_{3/2} - ^2P_{1/2}$    & 34.8                      &9.4                         &  8.1                         & $3.4 \times 10^5 $ [H], $1 \times 10^3 $ [e]     \\
	                     
	                     &\ArII\  $^2P_{1/2} - ^2P_{3/2}$    & 7.0                        & 2.0                         & 15.8                      & $4.0 \times 10^5 $ [e]     \\
	  \hline 
            MOPRA\tablefootmark{c}    &$^{12}$CO J = 1${\rightarrow}$0  & 2600                    & 43                          &  --                           & $1.8 \times 10^3 $ [\HH]      \\
	   \hline
	  
	  ASTE\tablefootmark{d}        &$^{12}$CO J = 3${\rightarrow}$2  & 867                         & 22                          &  --                           & $3.2 \times 10^4 $ [\HH]    \\
	  
	  \hline
	  
        \end{tabular}
        
       \tablefoot{
      \tablefoottext{a}{Critical densities are noted [e] for collisions with electrons ($T=10~000K$), [H] with hydrogen atoms ($T=100~K$) and [\HH] with molecular hydrogen ($T=10~K$, in the optically thin limit).}
      \tablefoottext{b}{\cite{Indebetouw2009}}
      \tablefoottext{c}{\cite{Wong2011}}
      \tablefoottext{d}{ \cite{Minamidani2011}}
      }
      \label{tab:properties}
  \end{table*}

  \subsection{ \her\ SPIRE spectroscopy}
  
    The SPIRE instrument includes an Imaging Fourier Transform Spectrometer (FTS) covering the wavelength ranges 194--324 and 316--672\mic\ 
   (SPIRE Short Wavelength SSW and SPIRE Long Wavelength SLW arrays respectively).
    30Dor was observed with the SPIRE FTS in the high spectral resolution ($\Delta \nu$ $\sim$ 1.2 GHz), intermediate spatial sampling mode.
    In the intermediate spatial sampling mode, SLW and SSW are moved between four jiggling positions  with a spacing of $\sim$28$''$ and $\sim$16$''$ respectively.
    The observations were performed on January 8, 2013 (observation IDs: 1342219550, 1342257932 and 1342262908) with a total integration time of $\sim15400$s.
        
We process the FTS data using the \textit{Herschel} Interactive Processing Environment (HIPE) version 11.0.2825 
and the SPIRE calibration version 11.0 \citep{Fulton2010, Swinyard2013}.
We use the method from \cite{Wu2013} to derive integrated intensity images and their uncertainties. 
This script has been recently used to generate FTS spectral cubes for M83 \citep{Wu2015}. 
A combination of parabola (continuum) and sinc (emission) functions is used to model a spectral line.
The spectra are then projected onto a grid that covers a 5$'$ $\times$ 5$'$ area with a pixel size of 15$''$ (roughly corresponding to the detector spacing for SSW).  
We perform a Monte Carlo simulation with 300 iterations to estimate the uncertainties on the spectra, as described in details in Lee et al (in prep).
The SNR is between 1 and 8 for \NII\ 205\mic\ and between 0.5 and 5 for \CI\ 370\mic.
The  \CI\ 609\mic\ is weaker and the SNR is below 2.

\begin{figure*}
     \centering
     \includegraphics[trim = 10mm 1mm 2mm 0mm, clip, width=6cm]{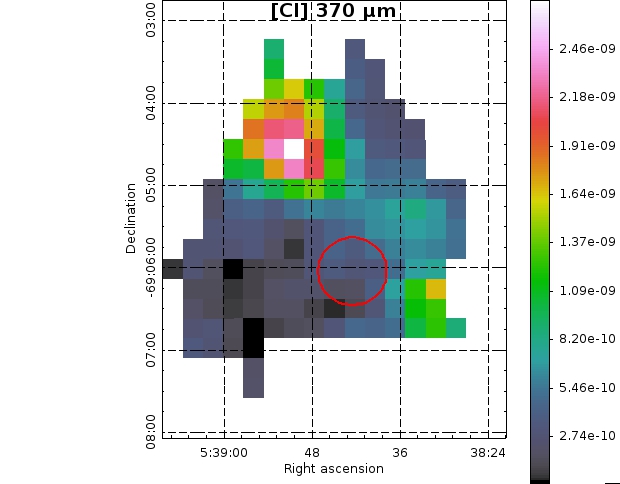}
     \includegraphics[trim = 10mm 1mm 2mm 0mm, clip, width=6cm]{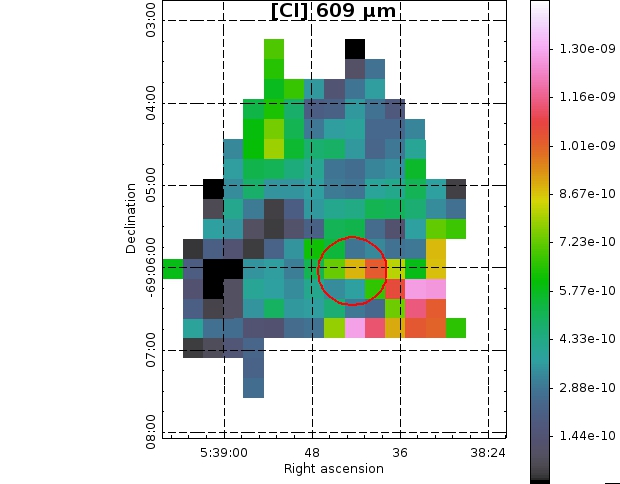}    
     \includegraphics[trim = 10mm 1mm 2mm 0mm, clip, width=6cm]{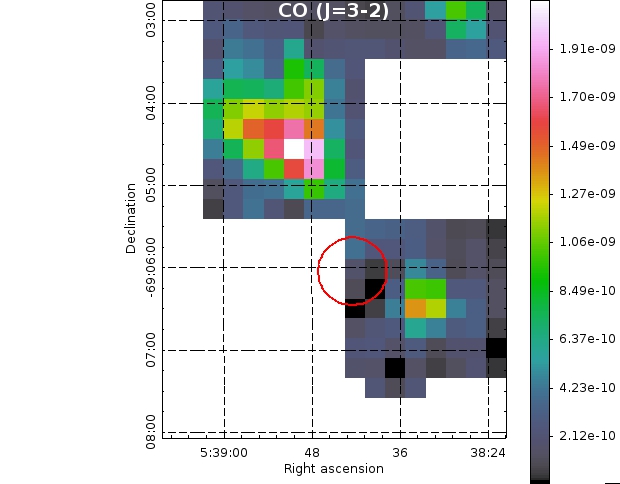} 
     \includegraphics[trim = 10mm 1mm 2mm 0mm, clip, width=6cm]{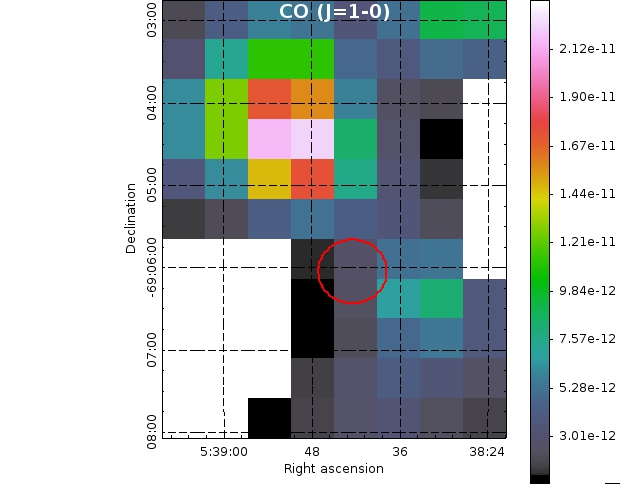} 
     \caption{\textit{Top panel}: \CI\ 370\mic\ (\textit{left}) and \CI\ 609\mic\ (\textit{right}) SPIRE/FTS maps of 30Dor.
     \textit{Bottom panel}: CO (3-2) ASTE map (\textit{left}) and CO (1-0) MOPRA map (\textit{right}). 
     All of the maps are shown at their original resolution, in W~m$^{-2}$~sr$^{-1}$.  
     The circle shows the location of R136.
     Table~\ref{tab:properties} and Section~\ref{Obs} describe these observations in detail.}
              \label{fig:CI-CO}
    \end{figure*}

The maps of \NII\ 205\mic\  and \CI\ 370 and 609\mic\ are presented in Figures~\ref{fig:maps} and~\ref{fig:CI-CO}.
Properties of these lines are presented in table~\ref{tab:properties}.
    CO transitions from J = 4 -- 3 to J = 13 -- 12 were also observed in 30Dor and will be presented in Lee et al. (in prep).

  \subsection{\her\ and \spit\ photometry}
    \label{sect:photometry}

To constrain the PDR models, we need to calculate the infrared luminosity, which requires photometry data from mid-infrared (MIR) to sub-millimeter.
PACS and SPIRE maps of the Large Magellanic Cloud at 100, 160 250, 350 and 500\mic\ were first published in \cite{Meixner2013} as part of the HERITAGE project.
We also use the observations of 30Dor obtained as part of the \spit\ \citep{Werner2004} Legacy program "Surveying the Agents of a Galaxy's Evolution" (SAGE ; \citealt{Meixner2006}).
We used the four channels of IRAC \citep{Fazio2004} at 3.6, 4.5, 5.8 and 8.0\mic\ and MIPS \citep{Rieke2004} observations at 24 and 70\mic .
The MIPS 24\mic\ map is saturated in several pixels.
We use the IRS spectra (\citealt{Indebetouw2009}; see also Sect.~\ref{sec:spitzer}) to calculate the 24\mic\ synthetic photometry in the MIPS 24 bandpass and compare to the original map, with excellent agreement in parts where the Spitzer/MIPS map is not saturated.
Table~\ref{tab:photometry} summarizes the photometry data we use to construct the infrared luminosity, associated with their spatial resolution.

  \begin{table}
       \centering
       \caption{Properties of the photometry data.}
       \begin{tabular}{l  c c }
	  \hline
	  \hline
	  Instrument  & $\lambda$ (\mic) & FWHM (arsec)  \\ 
	 \hline 
            IRAC\tablefootmark{a}    &3.6                    &     1.7    \\
                               & 4.5                    &      1.7   \\
                               & 5.8                    &      1.7   \\
                               & 8.0                    &       1.9  \\
	  \hline
	  MIPS\tablefootmark&24                   &      6      \\  
	  
	                      & 70                     &                18              \\
	                    % & 160                   &               36              \\
	   \hline
	  
	 PACS\tablefootmark{b}          & 100                      &             7.7         \\
	  
	                      &160                       & 12                        \\
	  \hline
	  SPIRE\tablefootmark{b}        & 250                     &        18               \\
	  
	                      & 350                     &             25            \\
	   
	                      & 500                      & 37                       \\
	  \hline

        \end{tabular}
        
       \tablefoot{
       \tablefoottext{a}{SAGE \citep{Meixner2006}}
       \tablefoottext{b}{HERITAGE \citep{Meixner2013}}
      }
      \label{tab:photometry}
  \end{table}

  \subsection{\spit\ / IRS spectroscopy}
    \label{sec:spitzer}

The Spitzer IRS low resolution data have been initially presented in \cite{Indebetouw2009}. 
The observed lines and their spatial resolution are listed in Table~\ref{tab:properties}.
We have reduced again the low spectral resolution cubes with CUBISM \citep{Smith2007} as part of an effort to measure lines that have not been investigated yet in detail, in particular the \HH\ lines, \SiII\ and \ArII .
The resolving power of both the short-wavelength/low-resolution (SL) and the long-wavelength/low-resolution (LL) modules range approximately from 60 to 120 (\spit\ Observer's Manual 7.1 2006\footnote{Available at http://ssc.spitzer.caltech.edu/documents/SOM.}).
The maps of  \SiII , \SIII\ 18\mic\ and \SIII\ 33\mic\ are presented in Figure~\ref{fig:maps}.

We use the total emission of the polycyclic aromatic hydrocarbon molecules (PAHs) that has been fitted by \cite{Indebetouw2009} using the package PAHFIT \citep{Smith2007a}.
Finally, we also use the Spitzer IRS high-resolution spectra presented by \cite{Lebouteiller2008} to measure the \HH\ lines.

    \subsection{Ground-based observations} 
  
  Low-J CO transitions are required for additional constraints for the PDR modeling. 
  We use the CO J = 1 -- 0 transition observed with MOPRA \citep{Wong2011} and CO J = 3 -- 2 observed with ASTE (\citealt{Minamidani2011}, see Figure~\ref{fig:CI-CO}).
  The spatial resolutions for the MOPRA and ASTE data are $42^{\prime \prime}$ and $22^{\prime \prime}$ respectively.
  
  The \Halpha\ emission, which we use as a qualitative tracer of the ionized gas, was observed with the Cerro Tololo Inter-American Observatory (CTIO) Curtis Schmidt telescope as part of the Magellanic Clouds Survey (MCELS, private communication; R.Leiton) at a resolution of $\sim5$\arcsec .

\subsection{Convolution kernels}
  
  As we use line ratios of different wavelengths and different instruments (see Table~\ref{tab:properties}), we must first convolve the maps to the same resolution.
  We add quadratically $12\%$ uncertainties to the PACS maps to account for the absolute calibration uncertainties (PACS Observer's Manual v2.5.1).
  When we use only PACS observations, all of the maps are convolved to the resolution of PACS at 160\mic\ ($12^{\prime \prime}$ or $\sim3$ pc), using the kernels from \cite{Aniano2011}.

  When we combine PACS, SPIRE and ground based spectroscopy data together to include the \CI\ and CO lines (see Sect.~\ref{sec:CI-CO}), all of the maps are smoothed to match the resolution of $42^{\prime \prime}$ ($\sim10$ pc), limited by the SPIRE long wavelength data.
  For this, we use the appropriate kernels
  to convolve a PACS point spread function (modeled by \citealt{Aniano2011}) to the SPIRE/FTS beam profile (fitted by a two-dimensional Hermite-Gaussian function) at the lowest resolution, 
essentially following the method by \cite{Gordon2008}.

The photometry bands are used to determine the infrared luminosity with our spectral energy distribution (SED) model (see Sect.~\ref{sec:IRmaps}).
Since we wish to perform the PDR analysis on the smallest possible scale (i.e. limited by the PSF of the PACS \CII\ map), we calculate the infrared luminosity at the resolution of $12^{\prime \prime}$, which is also the resolution of the PACS 160\mic\ band.
We have compared this approximated infrared luminosity to that determined using all available bands, i.e., including the SPIRE bands (250, 300 and 500\mic ). We find little difference on the integrated luminosity per surface area.
Thus we include only the bands between 24 and 160\mic\ to fit the SED at the best spatial resolution possible.

\subsection{Infrared luminosity maps}
\label{sec:IRmaps}
For each pixel of the map, we construct the full MIR to submm SED,
to which we apply the dust SED model of \citet[AC composition]{Galliano2011}.
This is a phenomenological SED fitting procedure with which we derive the resolved total infrared luminosity (\Ltir ) between 3 and 1000\mic , as well as the far-infrared luminosity (\Lfir ) between 60 and 200\mic. % by fitting the photometry data per pixel. % (\Lfir). 
This model was designed to fit the \her\ broadband photometry of the LMC, still remaining consistent with the elemental abundances.
The free parameters include the dust mass (from which the extinction magnitude in $V$ band, \Avdust , can be derived), the minimum starlight intensity, %$U_{min}$
 the difference between the maximum and minimum starlight intensities, %$\Delta U$
 the starlight intensity distribution power-law index %$\alpha_U$ 
and the PAH-to-total dust mass ratio $f_{PAH}$.
The final map of \Ltir , which integrates the SED fit between 3 and 1000\mic , can be seen in Figure~\ref{fig:maps}.
Contrary to other dust parameters, the infrared luminosity is very marginally model dependent.
It depends mainly on the wavelength coverage of the photometric constraints used, which in our case is sufficient.
For the PDR modeling (see Section~\ref{PDR}) we use the \Lfir\ integrated between 60 and 200\mic .
The SED model is better constrained in this limited wavelength range compared to the fit including longer wavelengths so this leads to less uncertainties on \Lfir .
However, even with no constraint longward 160\mic , the \Ltir\  of 30Dor is still relatively well constrained, as the dust in this region is sufficiently warm to peak at much shorter wavelengths.

  %__________________________________________________________________

\section{Data analysis}
  \label{Data}

   \subsection{General morphology}

   Figure~\ref{fig:maps} shows the maps of \Halpha, \OIII , \SIII , \NII , \CII , \SiII\ and \OI\ lines at their initial resolution.
   We also show the \Ltir\ (integrated between 3 and 1000\mic ) as well as the PAH emission (\Lpah ).
   The emission of all lines is distributed in the northern and southern lobes around R136.
   The emission of all lines peaks near the same location within 5~pc towards the northern lobe, with some inhomogeneous emission in the south.
   The spatial distributions of the \CI , CO(1-0) and CO(3-2) emission are presented in Figure~\ref{fig:CI-CO}.
   They show two lobes of emission as well, and the peaks of both the \CI\ and CO emission are shifted about 20\arcsec\ north of the peak of \CII .
   
   The \CII, \OI\ and \Lpah\ follow approximately each other throughout the map.
   The \OI\ 63\mic , \OI\ 145\mic\ and \CII\ emission lines, as well as \Lpah , are PDR tracers, although there may be a diffuse component as well, which could be contributing to the emission of \CII\ (see Section~\ref{sec:CIIemission}).
   
   The distributions of the \Halpha , \OIII\ 88\mic, \SIII\ and \NII\ lines have a structure different from the neutral PDR tracers \CII\ and \OI .
The spatial distributions of \OIII , \SIII\ and \NII\ follow well the distribution of the ionized gas traced by \Halpha\ emission and show in particular a characteristic arm-like structure in the north-east of R136.
   The ionization potentials of O$^{++}$ and S$^{++}$ are respectively 35 eV and 23.3 eV (see table~\ref{tab:properties}), so \OIII\ 88\mic\ and both \SIII\ 18\mic\ and \SIII\ 33\mic\ probe the highly ionized gas.
    The peak of \OIII\ is shifted from that of \CII\ toward the south, in the direction of R136.
    The peak of \NII\ 122\mic\ is located between the peaks of \OIII\ 88\mic\ and \OI , as expected from the values of the ionization potential of each specie. 
    \NII\ traces the low density and low-excitation ionized gas (critical density for collision with electrons are $\sim$310 cm$^{-3}$ and $\sim$48 cm$^{-3}$ for \NII\ 122\mic\ and \NII\ 205\mic\ respectively).
    
    The distributions of \SiII\ 35\mic\ and \Ltir\ emissions share properties both with  the ionized gas (\OIII\ and \SIII ) and the PDR tracers (\OI\ and \CII ).
    The \SiII\ 35\mic\ line and the \Ltir\ can in principle be used as a PDR tracer, but we also find in the maps some features that seem spatially associated with the ionized gas as well. 
    This is discussed in Sections~\ref{sec:CIIemission} and~\ref{sec:FIRemission}.
 \newline{}

\begin{figure}
      \centering
      \includegraphics[trim = 70mm 0mm 37mm 0mm, clip, width=8cm]{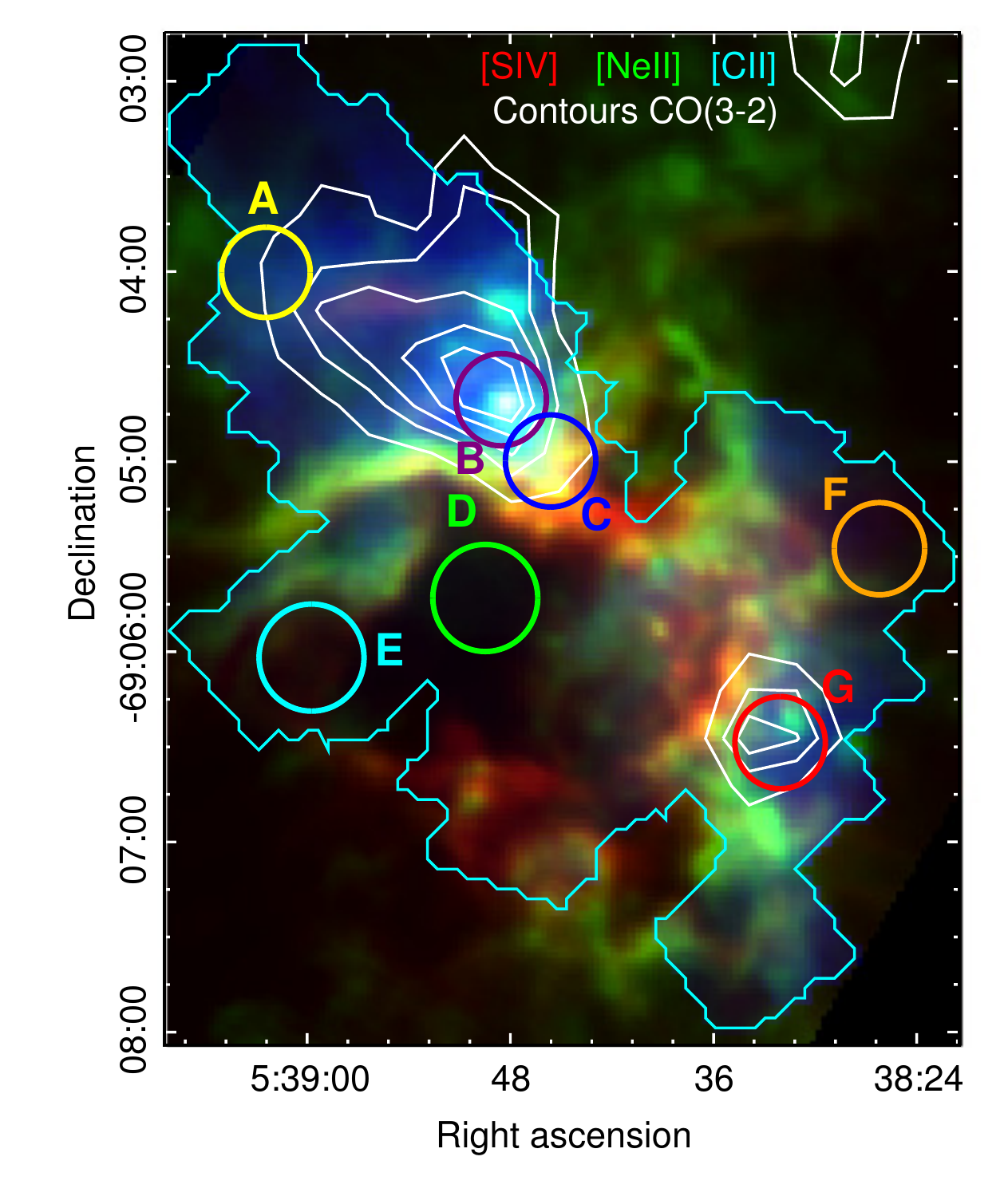}
      \caption{Red: Spitzer/IRS \SIV ~10.5\mic\  \citep{Indebetouw2009}. Green: Spitzer/IRS \NeII ~12.8\mic\  \citep{Indebetouw2009}. Blue: PACS \CII\ 158 \mic. 
      The blue contours outline the limit of the \CII\ map. We can identify the different layers of the gas from the highly ionized medium near the star cluster R136 to the clumpy molecular gas.
      In white are the contours of the $^{12}$CO(3-2) emission observed with ASTE \citep{Minamidani2011}. 
      The yellow region (A) is located on the north-east of R136, where \CII\ is extended but where \OIII\ seems fainter.
      The violet circle (B) is located on the \CII\ peak, the blue circle (C) on the \OIII\ peak.
      The green (D), cyan (E) and orange (F) circles probe more diffuse regions.
       The south-west peak of \CII\ is in the red region (G).
      } 
      \label{fig:3colors_2}
\end{figure}

 Figure \ref{fig:3colors_2} shows the different layers of the ISM, from the ionization front near the stellar cluster, where the highly ionized gas traced by the \SIV\ 10.5\mic\ emission is located, to lower ionization states (\NeII\ 15.6\mic ) and then to the PDRs traced by \CII .
The CO peak is located close to the \CII\ peak.
This spatial disposition suggests that R136 dominates the photoionization.
We can also see that the northern region seems to be quite well shielded from ionizing UV photons, since \CII\ is very extended in this direction while the tracers of the ionized gas show a sharp decrease on the other side of the \Halpha\ arc.
\newline{}

\begin{figure}
      \centering
      \includegraphics[trim = 2mm 10mm 2mm 10mm, clip, width=8.7cm]{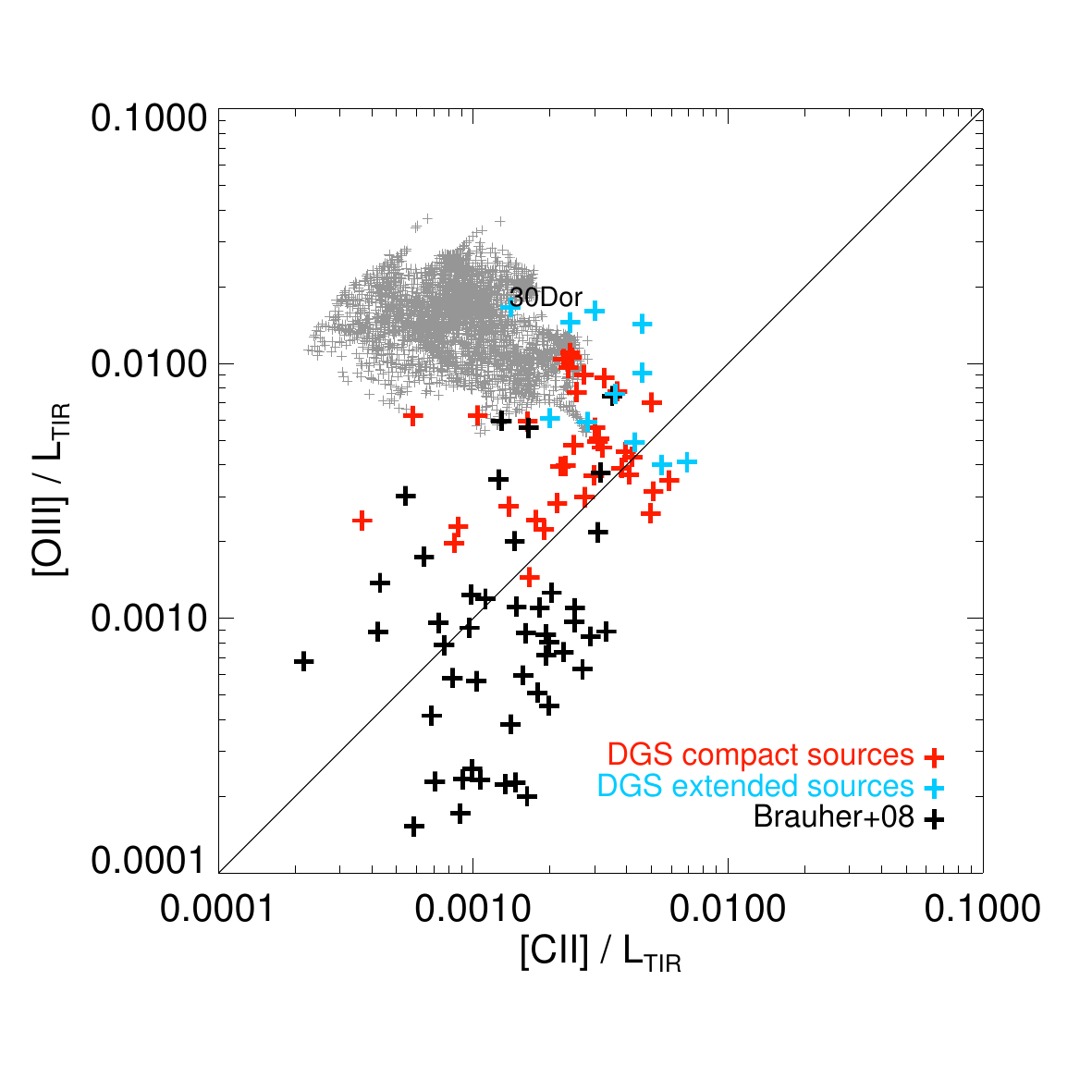}
      \caption{\OIII/\Ltir versus \CII/\Ltir. 
      The regions in 30Dor form the grey cloud with the high \OIII/\Ltir values.
      The DGS galaxies \citep{Cormier2015} are represented by red (compact) and cyan (extended) symbols. 
      The (mostly) metal-rich galaxies from \cite{Brauher2008} are represented by black symbols. 
      } 
      \label{fig:oiiiLtir_ciiLtir}
\end{figure}

  % In terms of brightness, 
  The \OIII\ 88\mic\ is the brightest FIR line in 30Dor, as in N11, the second largest \HII\ region of the LMC \citep{Lebouteiller2012}, the dwarf galaxy Haro 11 \citep{Cormier2012} 
   and in most of the dwarf galaxies \citep{Cormier2015}, integrated over full galaxy scales.
 This was first noted in several dwarf irregular galaxies in \cite{Hunter2001}.
 \OIII\ is brighter than \CII\ by a factor of 2 to 60 throughout our 30Dor map.
 Figure~\ref{fig:oiiiLtir_ciiLtir} shows the ratio \OIII/\Ltir\ versus \CII/\Ltir\  in 30Dor.
\cite{Cormier2015} already noted the elevated \OIII\ 88\mic /\Ltir\ in dwarf galaxies compared to the normal galaxies of \cite{Brauher2008}.
 Given that it requires 35~eV to ionize O$^+$ to O$^{++}$, this suggests the presence of high temperature stars throughout the region. 
 The range of \CII/\Ltir\ values covered in 30Dor is very broad (more than an order of magnitude) 
 and they cover almost the entire range of \CII/\Ltir\  observed in the wide range of galaxy type in Brauher et al. sample.
 The \OIII/\Ltir\ distribution is narrower (about a factor of 6 over the map).
 The regions with the highest \Ltir\ are the peaks of \OIII\ and \CII\ (Fig.~\ref{fig:3colors_2}, regions C and B). 
The highest \CII/\Ltir\ ratio is found in the northern part of 30Dor (near region A): this is due to the fact that \Ltir\ decreases more rapidly than \CII\ with increasing distance from the exciting sources. %\CII\ is very extended in the very porous region while \Ltir\ decreases rapidly as the distance from the exciting source increases.

 %%%%%%%      OIII/TIR vs CII/TIR

   In the regions where \CII\ is the brightest, the line intensities of \OI\ 63\mic\ and \SiII\ are similar to the \CII\ intensity.
 \OI\ 63\mic\ is everywhere at least ten times brighter than \OI\ 145\mic .
 The map of \NII\ 122\mic\ is the smallest and this line is also the faintest of our PACS lines (at least 200 times fainter than \OIII ), but the SNR throughout the region mapped is $\geq$5.
 The \NII\ 205\mic\ emission is 1 to 3 times fainter than \NII\ 122\mic .
   \newline{}

 \subsection{Origin of \CII\ and \SiII\ emission}
    \label{sec:CIIemission}
  
  Because the ionization potential of C$^0$, 11.3 eV, is lower than 13.6 eV, the \CII\ line can originate either from the PDRs or from the ionized gas.
  We thus need to investigate the possible contribution from the ionized gas and from PDRs to the \CII\ emission before we can use it as a PDR tracer and a constraint for PDR modeling.
  
\begin{figure}
     \centering
     \includegraphics[trim=10mm 0mm 0mm 0mm, clip, width=8cm]{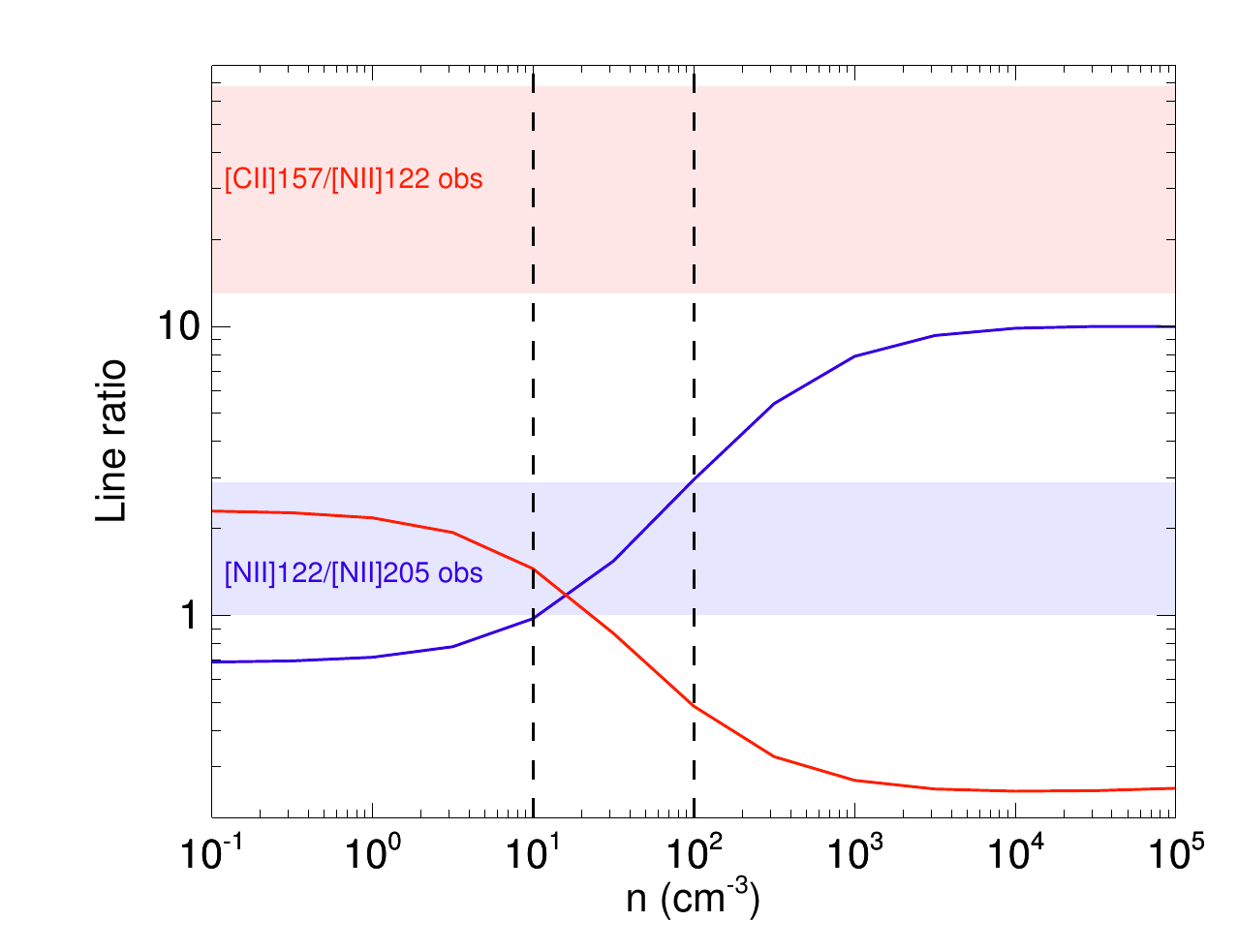}
     \caption{Theoretical ratios \NII~122\mic\ / \NII~205\mic\ (blue) and \CII\ / \NII~122\mic\ (red) at the temperature of 10 000 K. Blue and red areas indicate our observed ranges for the \NII~122\mic\ / \NII~205\mic\  and the \CII\ / \NII~122\mic\ ratios respectively. 
     The observed values of the \CII/\NII\ ratio (in red) are much higher than the theoretical value in the ionized gas for the entire map, indicating that \CII\ is mostly emitted in the PDRs and not in the ionized gas.
     }
              \label{fig:CIIth/NII}
    \end{figure}

\begin{figure}
     \centering
     \includegraphics[trim=38mm 5mm 10mm 0mm, clip, width=8cm]{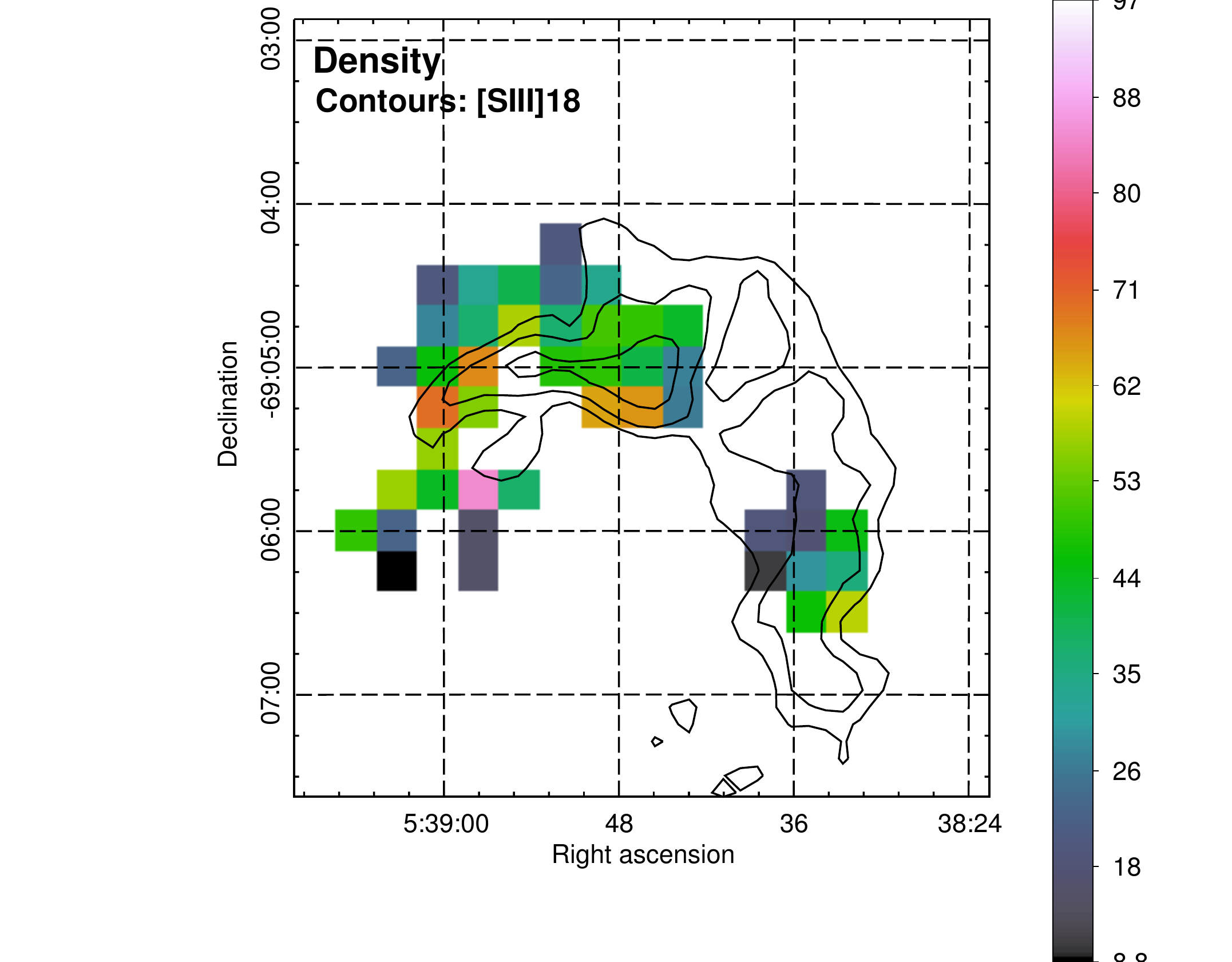}
     \caption{Electron density in cm$^{-3}$ calculated from the ratio \NII~122\mic\  / \NII~205\mic .  The pixel size is $20^{\prime\prime}$ ($\sim5 $pc). The black contours represent the emission of the ionized gas tracer \SIII\ 18\mic .}
              \label{fig:densNII}
    \end{figure}

  Following the analysis of \cite{Oberst2011}, we use the fact that the ratio \CII /\NII\ 122\mic\ can be calculated theoretically in the ionized gas, and that \NII\ originates only from the ionized gas.
  We calculate the fine-structure level populations of C$^+$ and N$^+$ as a function of the density using the theoretical collisional rates.
  We then apply a correction factor due to the ionic abundance fraction $ \displaystyle \frac{\text{C}^+/\text{C}}{\text{N}^+/\text{N}}$.
  We used the MAPPINGS III photoionization grids \citep{Allen2008} and found that this fraction depends little on the conditions (ionization parameter, starburst age, density) with a value around $0.85\pm0.15$.
  Finally, we scale the emission ratio with the observed elemental abundances of C and N in 30Dor from \cite{Pellegrini2011} listed in Table~\ref{tab:PDR}.
  The final ratio \CII /\NII\ 122\mic\ depends strongly on the electron density between 1 and 1000 cm$^{-3}$.
  Since the critical densities for \NII\ 122\mic\ and \NII\ 205\mic\ are 310 cm$^{-3}$ and 50 cm$^{-3}$ respectively (Table~\ref{tab:properties}), the ratio \NII\ 122\mic /\NII\ 205\mic\ is a good density tracer for the relatively low density ionized gas phase.
  We calculate the ratio \NII\ 122\mic /\NII\ 205\mic\ to determine the density using the theoretical curve from \cite{Bernard-Salas2012} (see Figure~\ref{fig:CIIth/NII}).
  This ratio depends only slightly on the temperature ; we choose a typical temperature of 10 000K.
  The calculated density presented in Figure~\ref{fig:densNII} ranges from 10 to 100 cm$^{-3}$. 
  Our values fall in the low density regime of the \SIII\ line ratio, which is sensitive to high density ($n_{crit} = 1.5 \times 10^4$ cm$^{-3}$ for \SIII\ 18\mic\ and $n_{crit} =4.1 \times 10^3$ cm$^{-3}$ for \SIII\ 33\mic ), and does not provide a useful constraint.

  If the \CII\ emission would originate in the low density ionized gas traced by \NII, the \CII /\NII\ 122\mic\ ratio would be $\sim0.5 - 1.3$  (Fig.~\ref{fig:CIIth/NII}).
  However, the observed ratio is significantly higher by a factor of $\sim 10$ than this theoretical ratio in the ionized gas. %, indicating that most of \CII\ is emitted in PDRs.
  At least $90\%$ of the \CII\ is expected to be emitted from PDRs in the entire mapped region. 
  The \CII\ emission can be considered to be a reliable tracer of the PDR gas in 30Dor.  
  
  Since the density is known and \ArII\ originates from the ionized gas, we can also calculate the theoretical ratio \CII/\ArII\  in the ionized gas. %, \ArII\ as a proxy for \CII\ in the ionized gas.
  Similarly, comparing the observed ratio \CII/\ArII\ to the theoretical ratio in the ionized gas, we also deduce that a large fraction (> 95\%) of the \CII\ emission originates from the PDRs.
  \newline{}
  
  \SiII\ emission can also originate from PDR or ionized gas.
  We proceed with the same method to separate the emission of the ionized gas from that of the neutral gas.
  The critical density of \SiII\ is close to that of \ArII\ ($3.4\times10^5$ cm$^{-3}$ and $4.0\times10^5$ cm$^{-3}$ respectively, see Table~\ref{tab:properties}) and both the observed ratios of \SiII/\ArII\ and \SiII/\NII\ are consistent with 60\% to 90\% of the \SiII\ emission originating from the PDRs.
  \newline{}

    \subsection{The photoelectric heating efficiency}
    \label{sec:PE}

%%%%%%%      CII/FIR vs FIR ,  OI63+CII/FIR vs FIR
\begin{figure}
      \centering
      \includegraphics[trim = 0mm 0mm 0mm 0mm, clip, width=8.7cm]{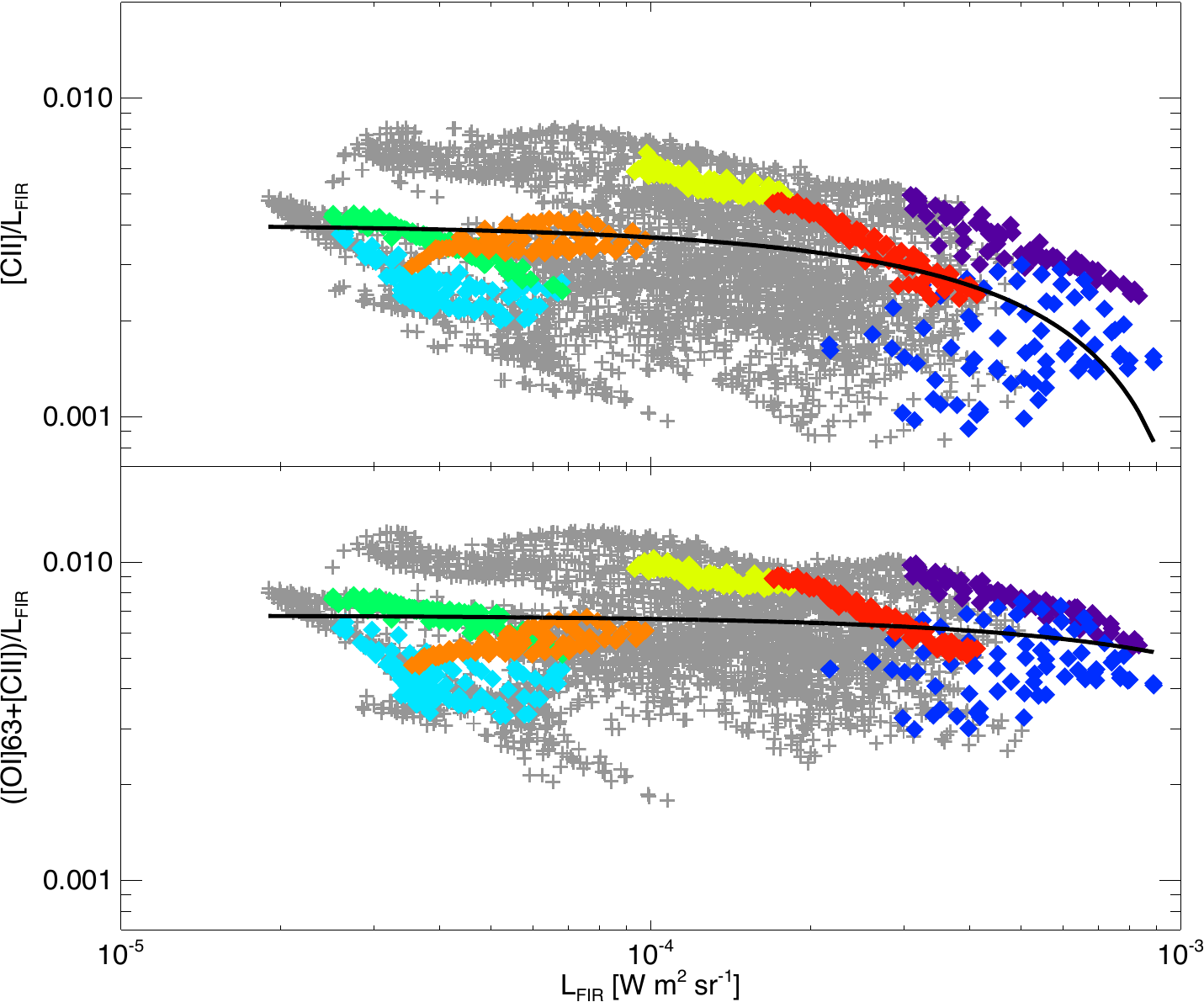}
      \caption{\textit{Top}: Ratio of \CII /\Lfir\ versus \Lfir . 
                     The color symbols are associated with the regions defined in figure~\ref{fig:3colors_2}. 
                     The grey symbols are all of the other pixels of our 30Dor map.
                     The black curve shows a linear regression of the data.
                      \textit{Bottom}: Ratio of (\OI\ 63\mic +\CII)/\Lfir versus \Lfir .} 
      \label{fig:ciiLtir_Ltir}
\end{figure}
The top panel of Figure~\ref{fig:ciiLtir_Ltir} shows the observed \CII /\Lfir\ ratio as a function of \Lfir\ for every pixel of the PACS map. 
The ratio \CII /\Lfir\ is often used to estimate the fraction of energy absorbed by dust that is used to heat the gas via the photoelectric effect (the photoelectric heating efficiency).
This ratio ranges between 0.1\% and 1\% with a significant scatter -- about one order of magnitude.
We observe a tendency of decreasing \CII /\Lfir\ as \Lfir\ increases (same trend as in \citealt{Stacey2010}), with a slope of $-3.6\pm0.2$. 
However, when we add the \OI ~63\mic\ emission (lower panel of Figure~\ref{fig:ciiLtir_Ltir}), we find a smaller dispersion (with a factor of 7) and also a flatter relation between (\OI ~63\mic + \CII ) / \Lfir\ and \Lfir , with a slope of $-1.4\pm0.3$ for the linear regression.
Both \OI\ and \CII\ are contributing noticeably to the cooling of the gas, as shown in \cite{Lebouteiller2012}.

%%%%%%%    OI+CII/FIR vs OIII/CII
\begin{figure}
      \centering
      \includegraphics[trim = 1mm 1mm 4mm 6mm, clip, width=8.7cm]{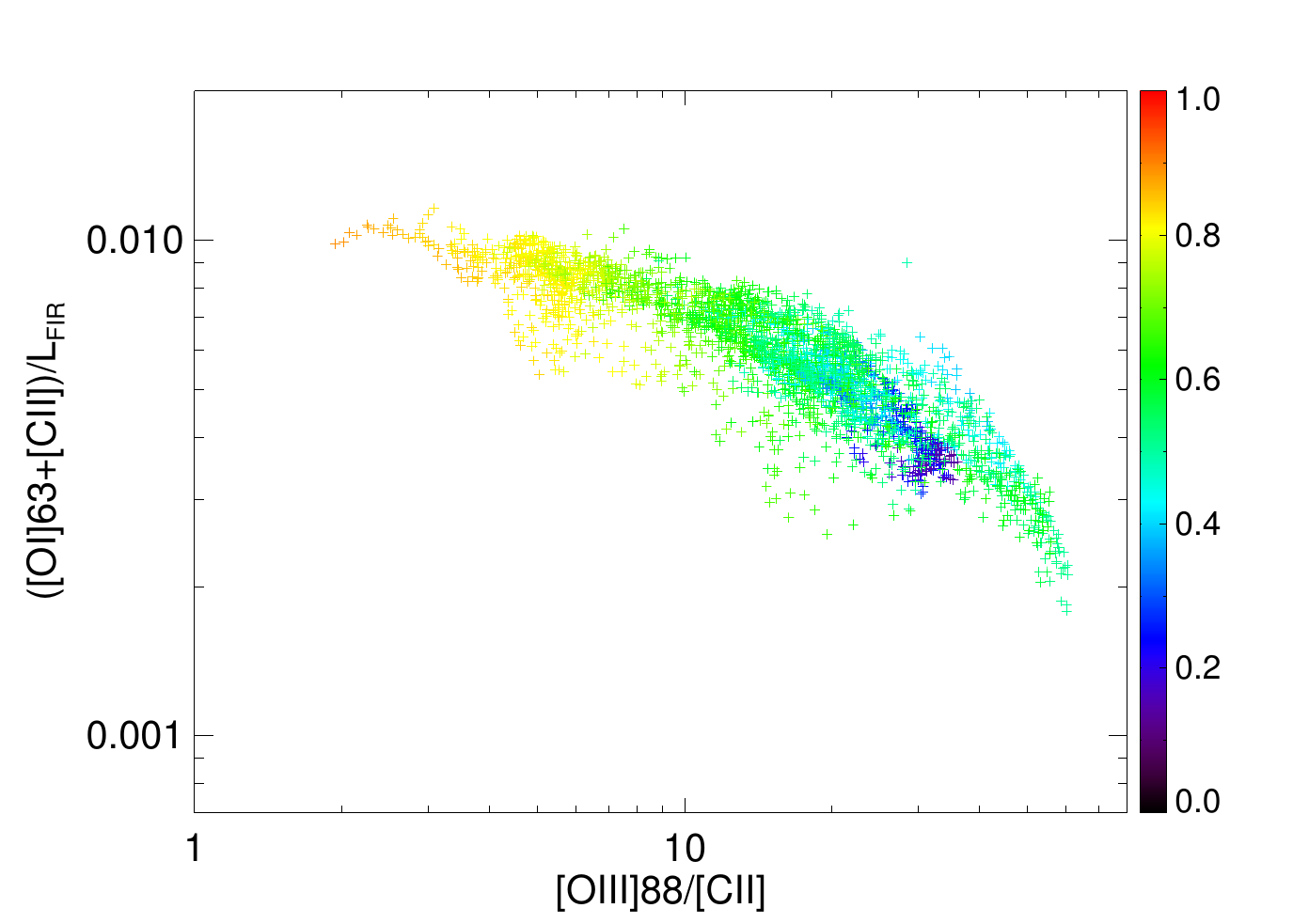}
      \caption{ Ratio of (\OI\ 63\mic +\CII) /\Lfir\ versus \OIII /\CII . The color bar indicates the fraction of \Lfir\ expected to come from the PDRs (see Section~\ref{sec:FIRemission}).} 
      \label{fig:oiciifir_oiiicii}
\end{figure}

Figure~\ref{fig:oiciifir_oiiicii} shows that the decrease of the ratio (\OI ~63\mic +\CII)/\Lfir\ is mostly due to the ionized component part in \Lfir .
Indeed, the ratio (\OI ~63\mic +\CII)/\Lfir\ is well correlated with the ratio \OIII\ 88\mic /\CII, and it decreases with increasing \OIII\ 88\mic /\CII, which is representative of the ionization state of the gas.
If we subtract the IR contribution from the ionized gas (as described in Section~\ref{sec:FIRemission}), we find that the ratio (\OI ~63\mic +\CII)/\Lfir\ to be fairly constant and narrow.
\newline{}

    \subsection{The origin of the FIR emission}
    \label{sec:FIRemission}
 
     \begin{figure*}[t]
     \centering
     \includegraphics[trim=0mm 0mm 0mm 0mm, clip, width=9cm]{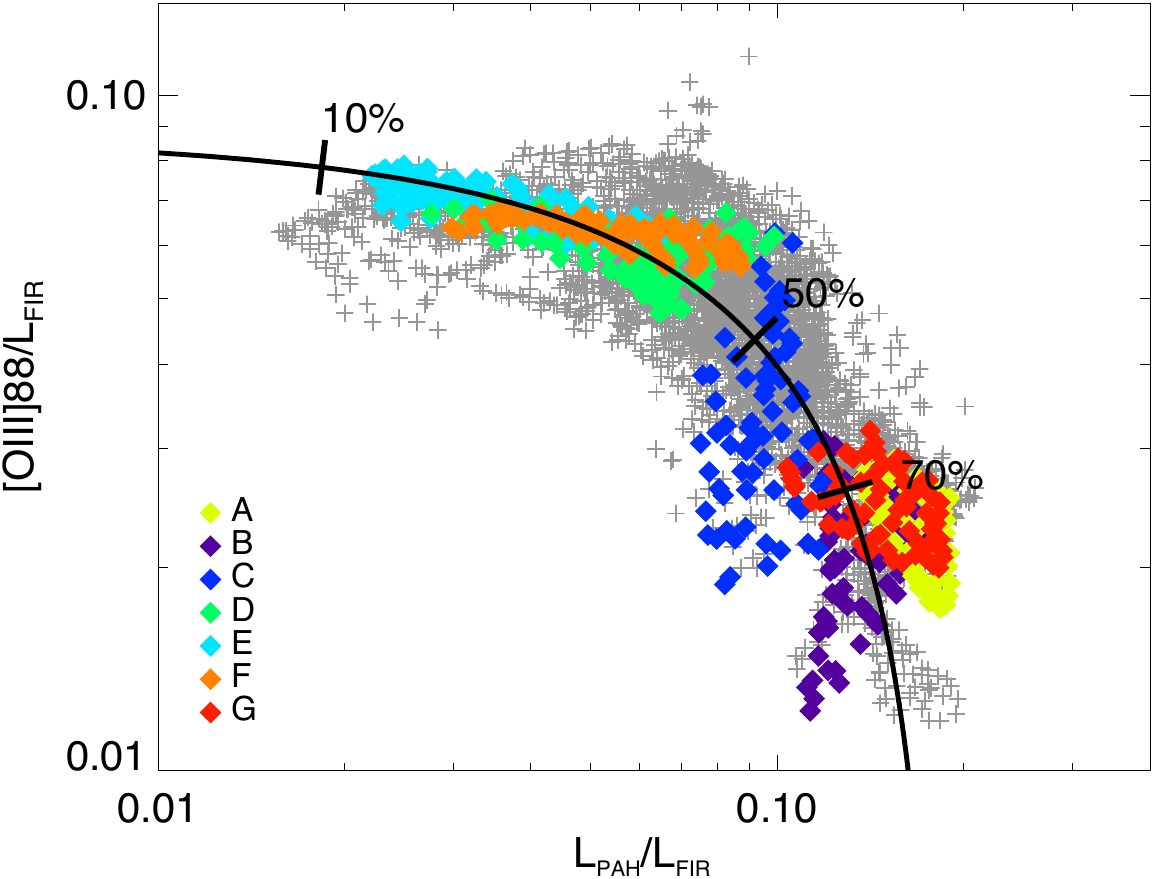}
    \includegraphics[trim=50mm 1.5mm 10mm 2mm, clip, width=8cm]{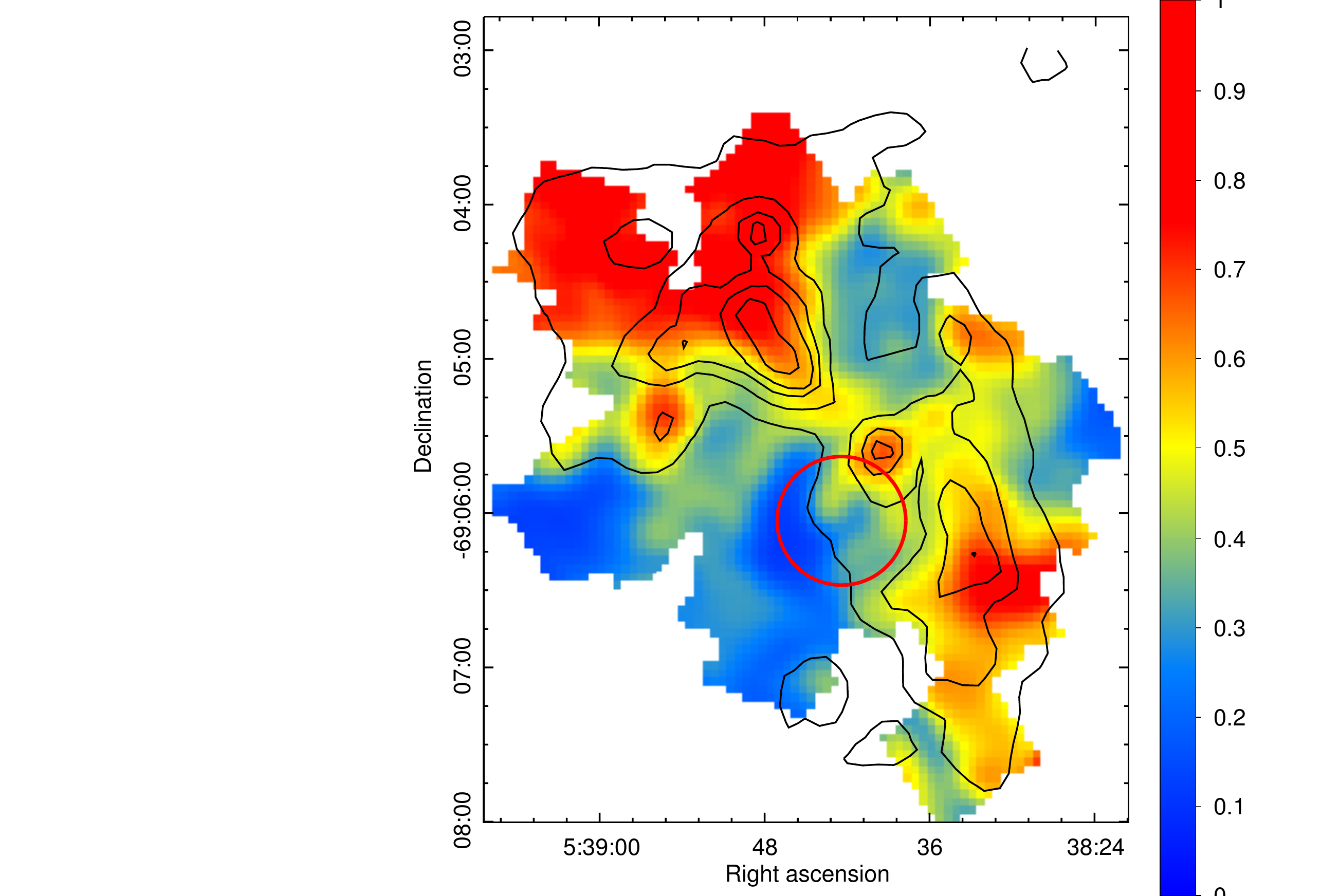}
     \caption{
     \textit{Left: }Ratio \OIII ~88\mic /\Lfir\ as a function of \Lpah /\Lfir .
                         The color symbols are associated with the regions defined in figure~\ref{fig:3colors_2}. 
                         The grey symbols are all of the other pixels of our 30Dor map.
                         The solid curve shows the result of the multiple linear regression, using the values $\alpha=5.4$ and $\beta=11.5$.
                         The values indicate the fraction of \Lfir\ coming from the PDR for 3 different positions on the curve.
    \textit{Right: } Ratio of \Lpdr /\Lfir . Black contours represent the total FIR emission. Close to the location of R136, only 20 to 30\% of the \Lfir\ come from the PDRs.
 }
              \label{fig:FIRdecomp}
    \end{figure*}

Although grains and PAHs in the PDRs contribute to the FIR emission, a fraction of \Lfir\ can also come from the \HII\ regions, from big grains in equilibrium with the interstellar radiation field (ISRF).
In this section, we inspect the origin of the \Lfir\ throughout the map, to separate the fraction in the ionized gas component from that in the PDR component.
In order to separate the emission from these two components, we assume that the PAH emission traces the PDRs and \OIII\ traces the ionized phase.
Indeed, we can see in Figure~\ref{fig:maps} that the spatial distribution of the \Lfir\ shows features similar to the PAH emission or any other neutral atomic gas tracer (\CII , \OI ), while other features seem to be spatially associated with \OIII\ or any other ionized gas tracers (\Halpha , \SIII ).
To disentangle the fraction of \Lfir\ in the ionized gas, we assume a linear relation such as: $ \displaystyle \text{\Lfir} = \alpha \times \text{\Lpah}$+$\beta \times \text{L}_{\text{\OIII}}$.
The pair of ($\alpha$~;~$\beta$) values is calculated using a multiple linear regression using all of the pixels in the map and is equal to (5.4 ; 11.5).
This decomposition implicitly assumes that the PAH-to-dust mass fraction is constant in PDRs and zero in \HII\ regions.
The left panel of Figure~\ref{fig:FIRdecomp} presents the correlation between \OIII ~88\mic /\Lfir\ and \Lpah /\Lfir .
The solid line on this plot shows the linear relation defined by:  $\displaystyle \frac{{\rm L}_{\text{\OIII}}}{\text{\Lfir}} = \frac{1}{\beta} -\frac{\alpha}{\beta}\times \frac{\text{\Lpah}}{\text{\Lfir}} $.
With this method, we are seeking a first order correction of the total \Lfir\ to be able to use it for the PDR modeling.
The modeled \Lfir\ reproduces the observed \Lfir\ within 30\% on average.
We determine the proportion of \Lfir\ coming from the PDRs as \Lpdr\ = $\alpha$ \Lpah .
The result is presented on the right panel of Figure~\ref{fig:FIRdecomp}.
On the north side of the map, far from the ionizing cluster, up to $90\%$ of the FIR emission is expected to come from the PDRs, while on the east of the map, near the R136 cluster, about 70\% of the FIR emission is expected to come from the ionized gas.
We subtract the estimated fraction of \Lfir\ emitted in the ionized gas, using \Lpdr\ for the PDR modeling.

    \subsection{Line ratios: empirical correlations} 

\begin{figure}
      \centering
      \includegraphics[trim = 7mm 3mm 5mm 5mm, clip, width=8cm]{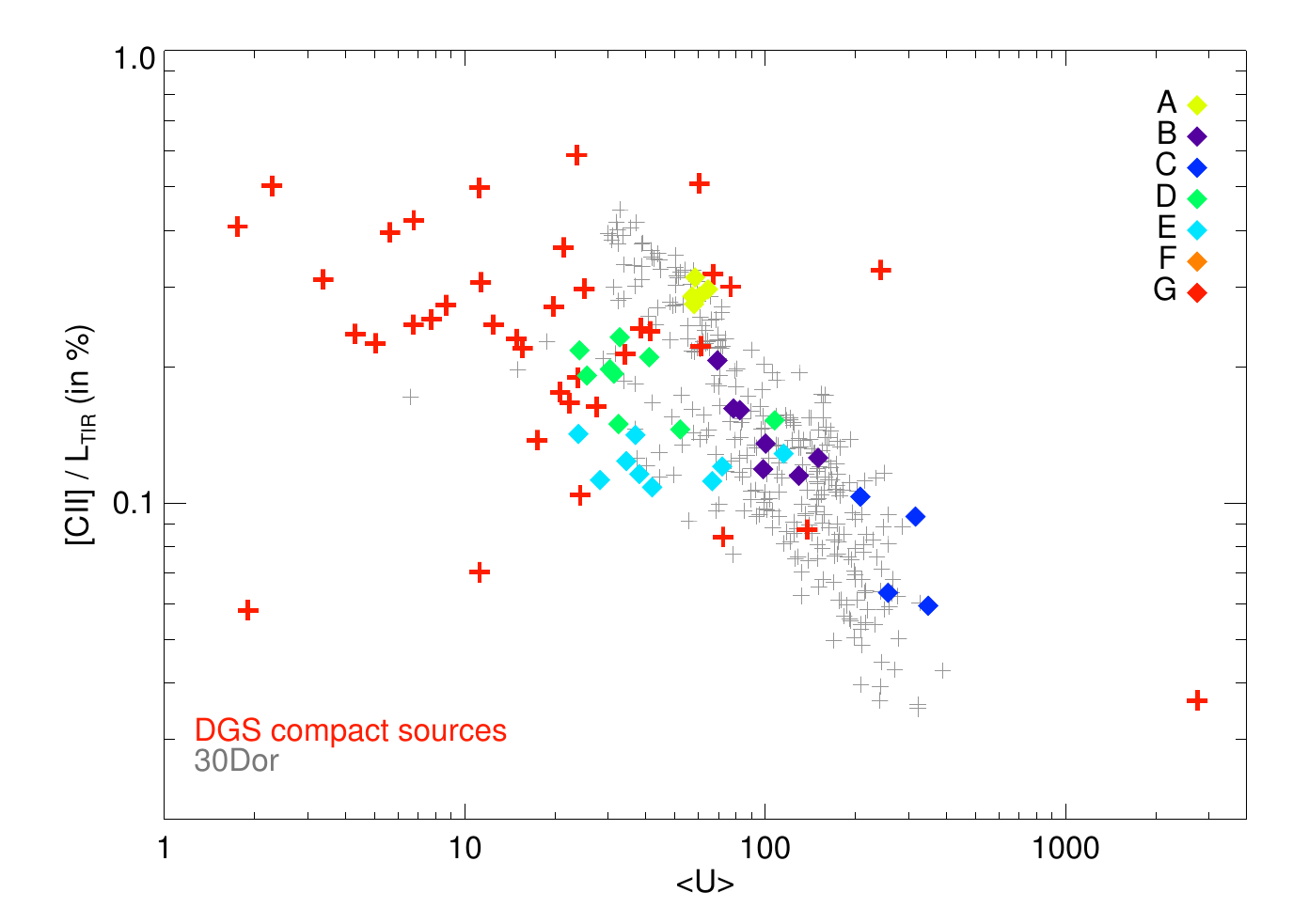}
      \caption{Correlation between \CII /\Ltir\ and <U> (average of the starlight intensity distribution). The red symbols are for the DGS galaxies \citep{Cormier2015}. The colored diamonds correspond to the regions of 30Dor illustrated in Figure~\ref{fig:3colors_2}} 
      \label{fig:Cooling_vs_Uav}
\end{figure}

\begin{figure}
      \centering
      \includegraphics[trim = 7mm 0mm 15mm 6mm, clip, width=8.7cm]{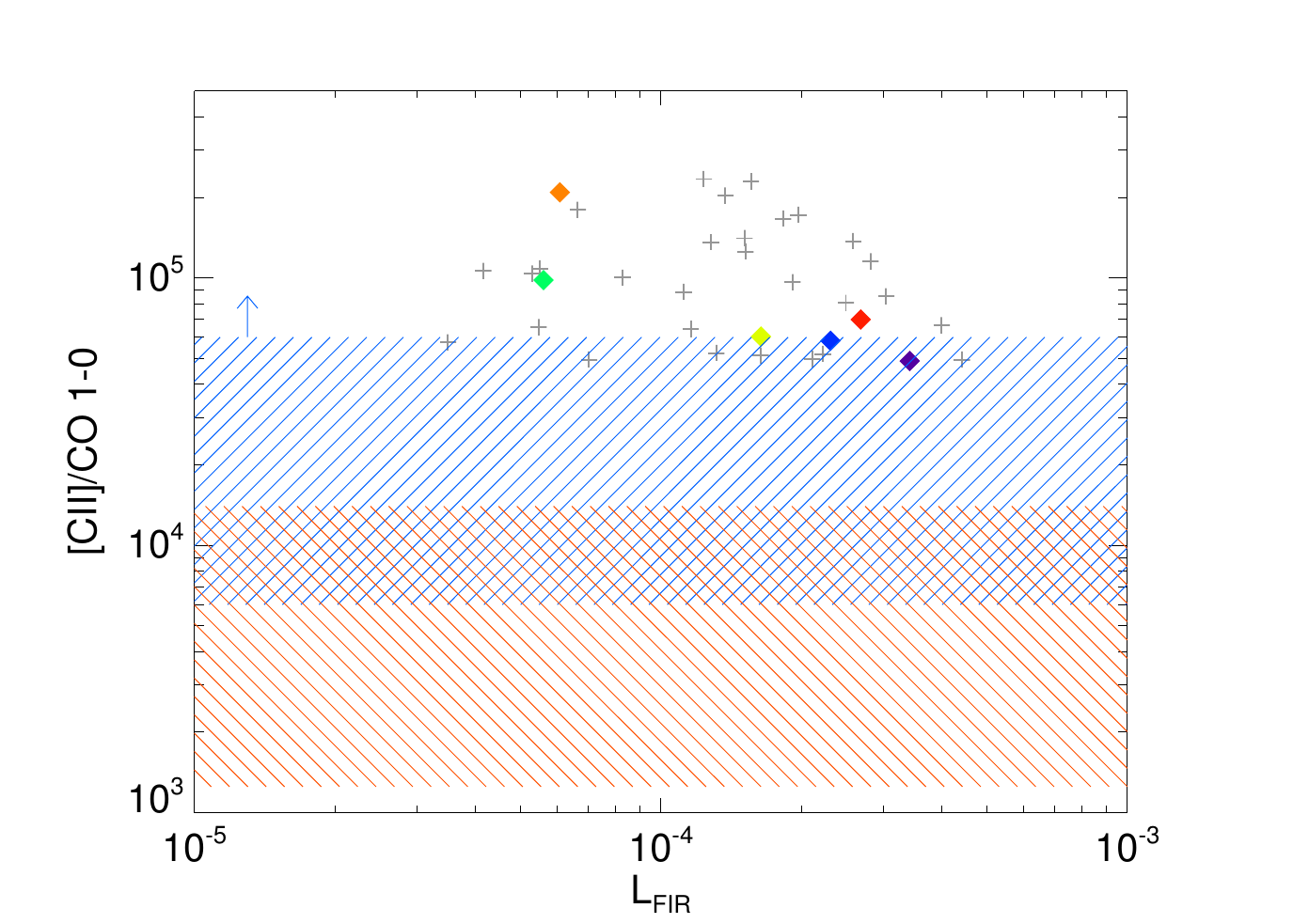}
      \caption{Ratio of \CII / CO(1-0) versus \Lfir\ at the scale of 30\arcsec . 
      The dashed horizontal orange band indicates the observed ratio \CII/C0(1-0) in star-forming regions of the Galaxy and metal-rich galaxies \citep{Stacey1991}.
      The dashed horizontal blue band indicates the range of observed ratios for 6 bright sub-solar metallicity galaxies of the DGS \citep{Cormier2014}.} 
      \label{fig:ciico_Ltir}
\end{figure}

Far-infrared line ratios are useful diagnostics of the ISM conditions.
We use these for PDR modeling (Section~\ref{PDR}) and we inspect here their distribution throughout 30Dor.
In Figures~\ref{fig:Cooling_vs_Uav} and \ref{fig:ciico_Ltir}), we focus into different regions of 30Dor to inspect the local variations.
\newline{}

%%%%%%%      CII/TIR vs <U>

To study how the distribution of starlight affects the observed photoelectric heating efficiency, we compare the average modeled starlight <U> from the SED modeling (see Sect.~\ref{sec:FIRemission}) with \CII /\Ltir\ (Fig.~\ref{fig:Cooling_vs_Uav}).
Note that we use the total \Ltir\ here since we do not know the fraction of the PDR component for the DGS sources.
We compare our spatially resolved values with the distribution of <U> and \CII/\Ltir\ from the integrated DGS compact sources of \cite{Cormier2015} and \cite{Remy-Ruyer2014}. % and 10\arcsec\ pixels in the 30Dor map.
More details and references can be found in \cite{Cormier2015}.
We see that 30Dor covers a large range in \CII/\Ltir , approximately one order of magnitude, and approximately one order of magnitude in <U>.

There is a trend of decreasing \CII/\Ltir\ as <U> is increasing, following the trend observed by \cite{Cormier2015}, showing an apparent line deficit at high <U> values. 
This is probably an effect of the contribution of the ionized component to the infrared luminosity, as shown in Figure~\ref{fig:oiciifir_oiiicii}.
The \OIII\ peak (region C in Figure~\ref{fig:3colors_2}) is the region with the highest <U>: it is the closest region to R136 in physical distance (see~\ref{sec:3D_geo}) ; the gas is mostly ionized.
This region corresponds also to the lowest \CII/\Ltir\ ratio.
\newline{}

%%%%%%%      CII/CO 1-0 vs FIR

To compare one of the primary PDR coolants with the CO$(1-0)$, we show in Figure~\ref{fig:ciico_Ltir} \CII /CO versus \Lfir\ at the resolution of 42\arcsec .
\CII/CO(1--0) throughout 30Dor is $\sim 5\times10^4 - 3\times10^5$. 
The observed range of the ratio \CII/CO(1-0) is very broad. %and \CII/CO 3-2 
These values are about a factor 10 higher than the typical values for spiral or starburst galaxies ($\sim 2000-8000$; \citealt{Stacey1991, Negishi2001}), 
and more in agreement with the range of values measured for integrated dwarf galaxies as already noticed in \cite{Poglitsch1995},  \cite{Madden1997} and \cite{Cormier2010, Cormier2014}.   
     We note that low metallicity galaxies always show extreme \CII /CO compared to the more metal-rich galaxies.

  %__________________________________________________________________

\section{PDR modeling}
  \label{PDR}
  
  In this section, we model the infrared observations of 30Dor using the Meudon PDR code and present results describing the properties of the gas.
  
  \subsection{The Meudon PDR model}  

The Meudon PDR code\footnote{The Meudon PDR code is public and available online at the following address: http://ism.obspm.fr.}
 is described in \cite{LePetit2006}, \cite{LeBourlot2012} and \cite{Bron2014}. It computes the atomic and molecular structure of interstellar clouds.
The model considers a 1D stationary plane parallel slab of gas and dust illuminated by a radiation field (from UV to radio) arising from one or both sides. 
The radiative transfer is solved in an iterative way at each point of the cloud by taking into account absorption by gas and dust and scattering and emission by dust.
For the present work, we used a development version (v1.6.0) with updates that includes the computation of X-ray 
radiative transfer and the impact on the chemistry and thermal balance of the cloud
(Godard et al. in prep).

   \subsection{Input and output parameters}
   
   We describe here some of the configuration parameters of the model.
   We assume that the gas in each pixel can be modeled by a single cloud of pressure P, illuminated by a radiation field. % \go .
   The standard radiation field used in the Meudon PDR code is that observed in the solar neighborhood \citep{Mathis1983} and is scaled with the parameter \go\ to control the intensity of the incident radiation field on each side of the cloud. 
   For \go ~=~1, the integrated energy density between 911.8~\AA\ to 2400~\AA\ is $6.8\times10^{-14}$ erg cm$^{-3}$.
In the model we ran, \go\ ranges between $1$ and $10^5$ on one side and is fixed to $1$ on the other side to expose this side to the general interstellar field. 
The pressure ranges between $10^4$ and $10^8$ cm$^{-3}$ K. 
The pressure is constant throughout the cloud, however a constant density model has also been explored (Sect.~\ref{sec:isobar}).
We have investigated the possibility of adding X-rays in the model, but found that they are not needed to explain the PACS observed data (Sect.~\ref{sec:excitation}).
The visual extinction of the entire cloud has been varied from $A_{\rm{V}}^{\rm {max}} = 1$ magnitude to $A_{\rm{V}}^{\rm {max}}  = 10$ magnitude.
We estimate the mass fraction of PAHs from our SED modeling (Sect.~\ref{sec:IRmaps}).
We find that $f_{\rm{PAH}}=1\%$ is adapted for 30Dor.
We use %a mass fraction of PAHs $f_{\rm{PAH}}=1\%$ adapted for 30Dor \citep{Galliano2011} and 
elemental abundances as in \cite{Pellegrini2011} for He, C, N, O, Ne, Si and S, to reproduce as accurately as possible the conditions in 30Dor (see table~\ref{tab:PDR}).
The dust-to-gas mass ratio is fixed to $0.5\times10^{-3}$ based on our SED modeling (Sect.~\ref{sec:IRmaps}).

  \begin{table}
       \centering
          \caption{Input parameters for the PDR model.}
       \begin{tabular}{l l c}
	  \hline
	  \hline
	  Parameter & Notation & Value                            \\ 
	  \hline
	  
	  Pressure    & P       &   $10^4 - 10^8$ cm$^{-3}$ K \\
	 Radiation field  &  \go\             &   $1 - 10^5 $ \\    %(scale to \citealt{Mathis1983})
	   Cosmic ray flux   &$\zeta$\tablefootmark{a}      &   $3\times10^{-16}$ s$^{-1}$\\
	  Total visual extinction &$A_{\rm V}^{\rm {max}}$         &   $1 - 10$ mag \\
	   Metallicity & $Z$\tablefootmark{b}             &   $0.5$ Z$_{\odot}$ \\
	   PAH fraction  &$f_{\rm{PAH}}$       &   1 \% \\
	  Dust-to-gas mass ratio & $M_{\rm{dust}}$/$M_{\rm{gas}}$ &  $5\times10^{^{-3}}$ \\
	  \hline
	  Gas phase abundances\tablefootmark{c}  &  &  $log(n(\text{X})/n(\text{H}))$\\
	  \hline
	  He            &   &    -1.05\\
	  C               &  &    -4.3\\
	  N               &  &    -4.91\\
	  O               &  &    -3.75\\
	  Ne             &  &    -4.36\\
	  Si              &  &     -5.51\\
	  S               &  &     -5.32\\
	 % Cl                &     -7.16\\
	  %Ar                &     -6.04\\
	  %Fe               &     -5.95\\
	  \hline
        \end{tabular}
        \tablefoot{
      \tablefoottext{a}{\cite{Indriolo2012, Indriolo2015}}
      \tablefoottext{b}{Metallicity: 12+log(O/H) = 8.38 \citep{Rolleston2002, Pagel2003} and (O/H)$_{\odot} = 4.9\times10^{-4}$ \citep{Asplund2009}.  }
      \tablefoottext{c}{\cite{Pellegrini2011}}
      }
      \label{tab:PDR}
  \end{table}

\begin{figure}
     \centering
        \includegraphics[trim=0mm 0mm 0mm 0mm, clip,width=8cm]{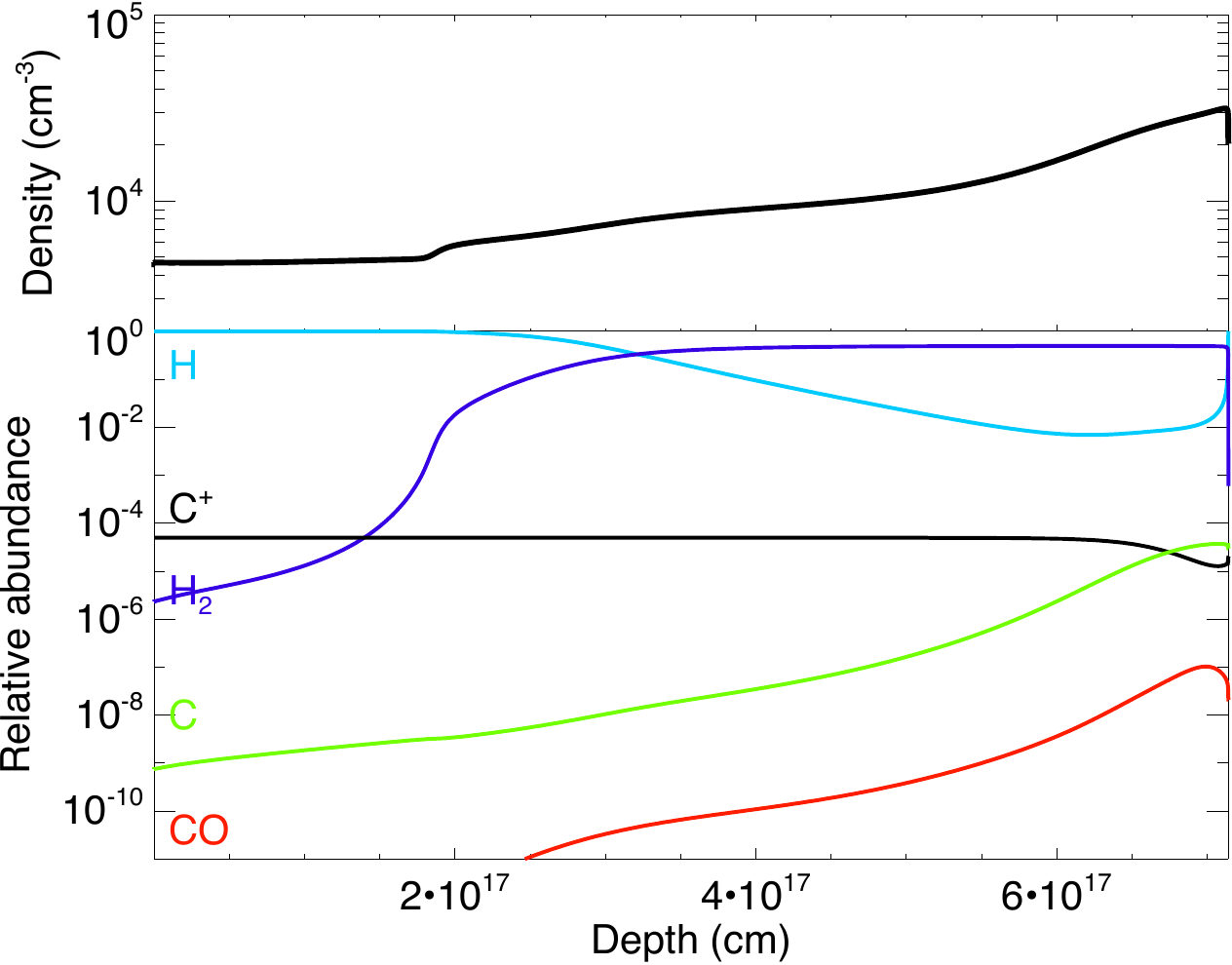}
     \caption{\textit{Top:} Density profile as a function of the depth into the cloud in a simulated cloud of \Avmax $=2$, 
                                          P $=10^6$ cm$^{-3}$ K with \go\ = 3000 on the left side and \go = 1 on the right side.
                     \textit{Bottom:} Local gas phase relative abundances of C$^+$, C, CO, O and \HH\  for the same model.
                      }
              \label{fig:abundance_profiles}
    \end{figure}

The output quantities computed by the model include the integrated line intensities and \Lfir , 
the ionic and molecular abundances, the emissivities and the chemical and thermal structure of the cloud.
An example of the variations of the local gas phase abundances for some elements as a function of the depth into the cloud is presented in Figure~\ref{fig:abundance_profiles} for a typical cloud of \Avmax\ of 10, with a constant pressure of P~$=10^6$~cm$^{-3}$~K, illuminated by a radiation field of \go\ = 3000.

       \begin{figure}
     \centering
      \includegraphics[trim=0mm 0mm 0mm 0mm, clip, width=9cm]{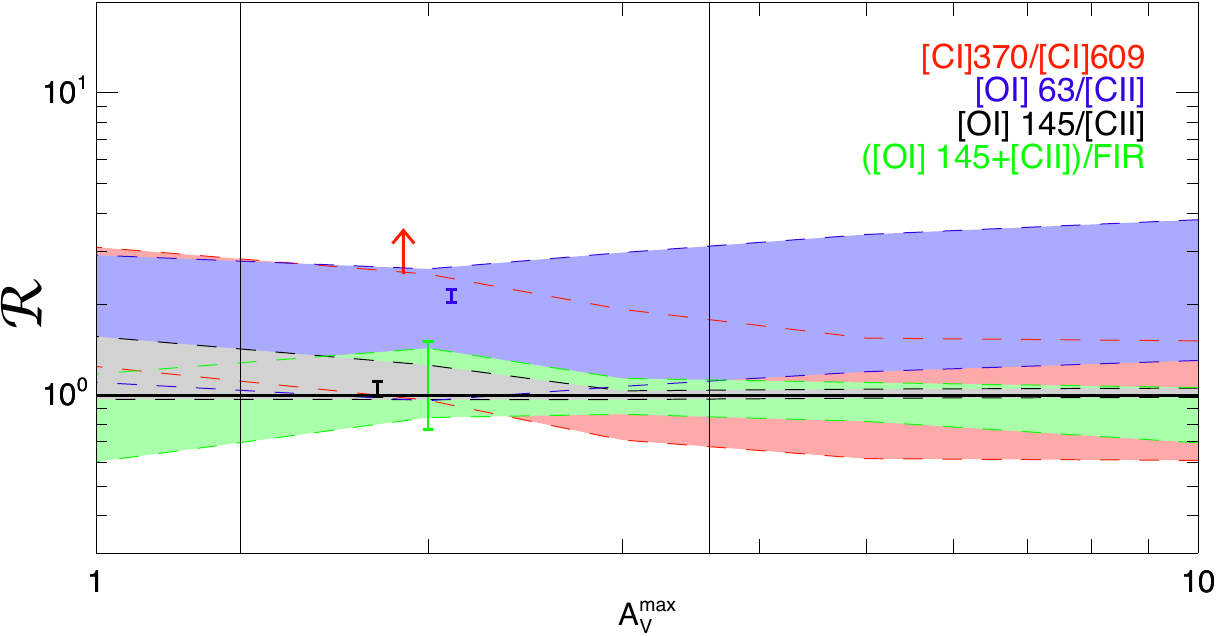}
      \caption{ Ratios, $\mathcal{R}$, of the modeled line ratios over the observed line ratios for \CI ~370\mic/\CI ~609\mic , \OI ~63\mic/\CII, \OI ~145\mic/\CII\ and (\OI ~145\mic+\CII)/\Lpdr , for simulated clouds of different \Avmax\ from 1 to 10, for 30Dor abundances.
      The width of the bands represents the dispersion of $\mathcal{R}$ throughout the map.
      The error bars due to uncertainties on the observations are plotted.
      The vertical lines indicate the range of \Avmax\ where the predictions of the model are compatible with the observed \CII / CO(3-2) (see Fig.~\ref{fig:30dor_Av}).
      }
              \label{fig:Av_dependance}
   \end{figure}

We use \OI, \CII\ and \Lpdr\ to constrain P and \go\ in the PDR model pixel by pixel.
We consider here the observed \CII\ emission without correction since the contribution from the ionized gas is low (Sect.~\ref{sec:CIIemission}) %(originating almost exclusively from the PDR), 
and we correct the \Lfir\ emission to remove the contamination from the component associated with the ionized gas (see Sections~\ref{sec:CIIemission} and~\ref{sec:FIRemission}).
At first we let \Avmax\ vary freely as this set of tracers cannot constrain this parameter and it has no influence on the resulting best model.
This is illustrated in Figure~\ref{fig:Av_dependance}, where we show the ratios $\mathcal{R}$ between several modeled line ratios and their observations for simulated clouds of different total depths. 
The ratios \OI /\CII , (\OI +\CII )/\Lfir\ are not very sensitive to \Avmax , as is also the case for the ratio \CI ~609\mic /\CI ~370\mic .

    \subsection{Pressure and incident radiation field}
    \label{sec:results}

We find the best solution for the incident radiation field and the pressure using the observed line intensities as constraints.
They are found by minimizing the $\chi ^2$ distribution : 
\begin{equation}
  \chi ^2 = \sum_{j=1}^{N} \left ({\frac{I_j(x,y)-M_j}{\sigma_j(x,y)}} \right )^2,
\end{equation}

where $I_j(x,y)$ is the observed value of the ratio $i$, for a given pixel (x,y), $\sigma_j(x,y)$ is the uncertainty associated with this observed ratio and $M_j$ is the value predicted by the model for the ratio $j$. $N$ is the number of constraints (independent ratios) that are used.
Ratios of line intensities are used instead of absolute values. % to be more conservative at first. 
In this case, if species are co-spatial in the cloud and if there are no opacity effects, 
we can then ignore the effect of an area filling factor different than one and the presence of several clouds along the line of sight (see Sect. \ref{sec:fill_factor}).

\begin{figure*}
     \centering
      \includegraphics[trim=0mm 0mm 0mm 0mm, clip,width=8cm]{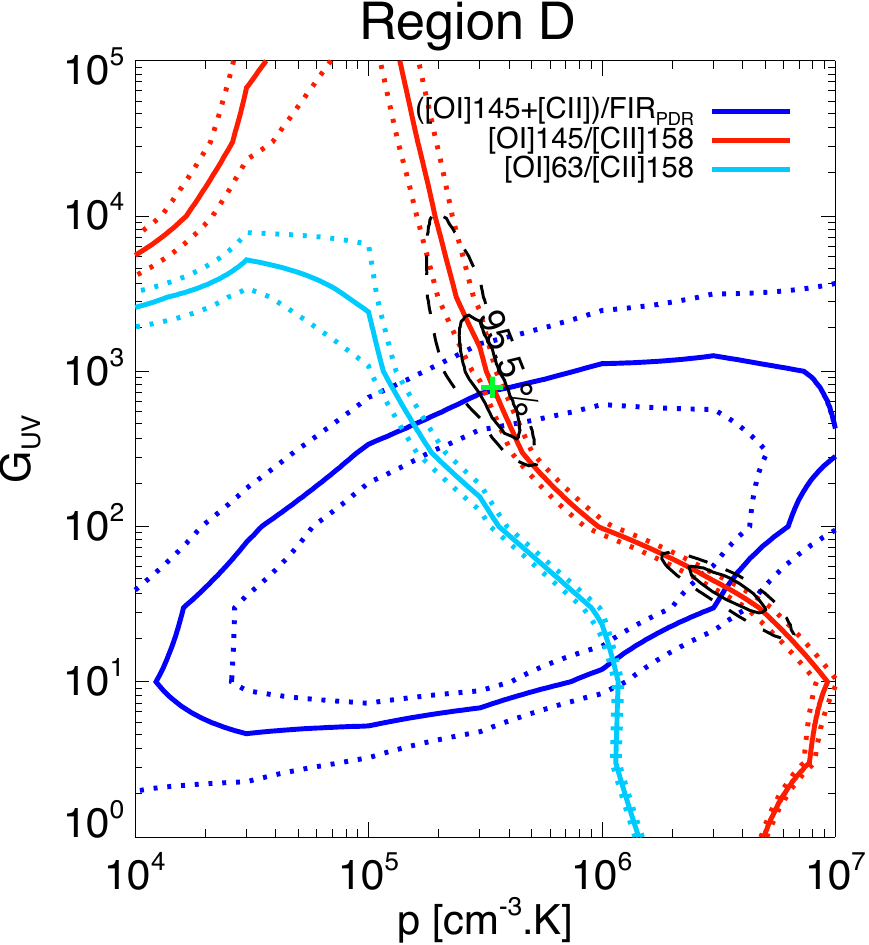}
      \includegraphics[trim=0mm 0mm 0mm 0mm, clip,width=8cm]{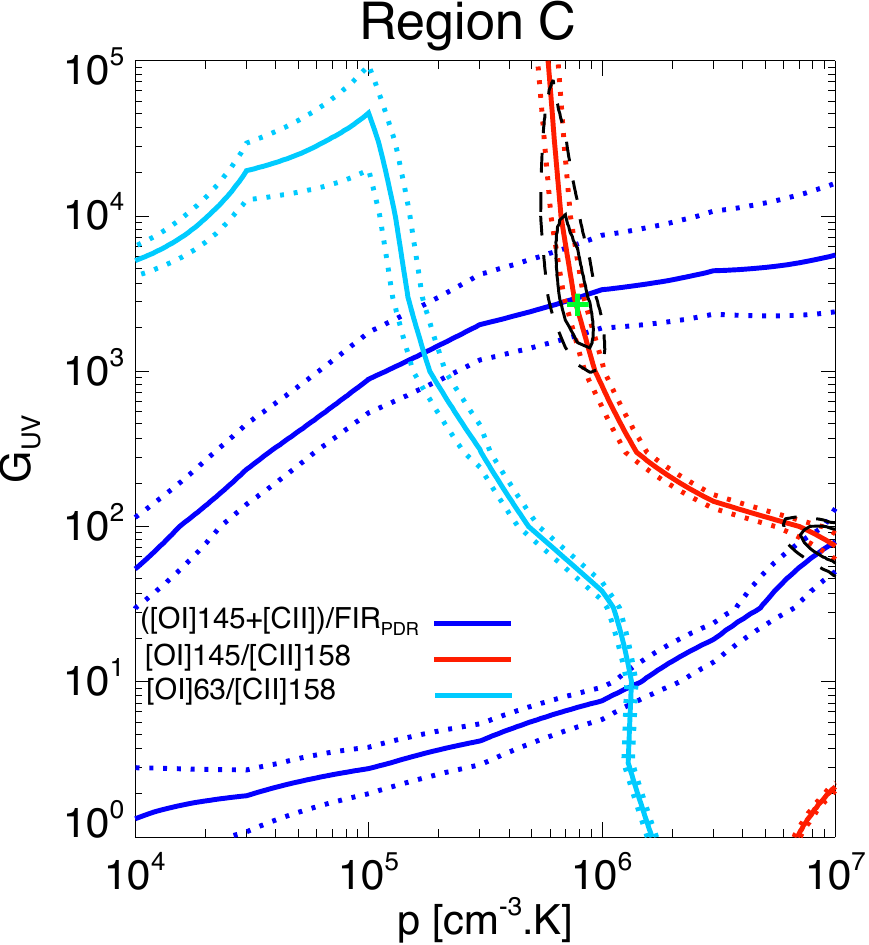}
     \caption{Contour plots showing the parameters \go\ and P from our model, which reproduces the observed values of the ratios (\OI ~145\mic+\CII)/\Lpdr\ (dark blue), \OI ~145\mic/\CII\ (red) and \OI ~63\mic/\CII\  (cyan). %and $\frac{\text{\CI}609}{\text{\CI}370}$ (green)
                      The dotted lines show the 1$\sigma$ error on the observed ratios.
                      The solid and dashed black contours represent the confidence intervals for, respectively, 1 and 2 $\sigma$. %and dotted, 3
                      The green crosses indicate the best solutions for \go\ and P in each case, using (\OI ~145\mic+\CII)/\Lpdr\ and \OI ~145\mic/\CII\ as model constraints.
                      Left plot is for a pixel representative of the region D (from Figure~\ref{fig:3colors_2}). Right plot is for a pixel from region C (from Figure~\ref{fig:3colors_2}), near the \go\ peak.}
              \label{fig:pixel_plot}
    \end{figure*}

As an example, we present in Figure \ref{fig:pixel_plot} the values of \go\ and P that reproduce the observed values for the ratios (\OI ~145\mic+\CII)/\Lpdr\ in blue, \OI ~145\mic/\CII\ in red and \OI ~63\mic/\CII\ in cyan for two %$ 3^{\prime \prime} \times 3^{\prime \prime}$ 
pixels of the map of 30Dor at $12 ^{\prime \prime}$ resolution, located in the regions D and C (Fig.~\ref{fig:3colors_2}).
We can see in these figures that the constraint given by the ratio \OI ~63\mic/\CII\ is never consistent with the other ratios within the error bars. 
This is likely due to optical depth effects in the \OI ~63\mic\ line, as we show later in this section.
For now, we will not consider this line to constrain the parameters of the model.

First, we use the ratios (\OI ~145\mic+\CII)/\Lpdr\ and \OI ~145\mic/\CII\ to constrain \go\ and P.
Thus we are limited by the PACS resolution of $12^{\prime\prime}$ and by the spatial coverage of the \OI\ 145\mic\ map.
We do not use the ratio \CI ~370\mic/\CI ~609\mic,
which does not bring strong constraints on \go\ and P as the error bars are very large.
We have then the same number of constraints and parameters.
We can note from Figure~\ref{fig:pixel_plot} that there is a degeneracy between a high \go /low P solution and a low \go /high P solution.
The addition of the ratio \CII /\CI\ or \CII /CO does not help to break this degeneracy since they are very dependent on \Avmax\ (Sect.~\ref{sec:CI-CO}).
However, the high \go /low P solution, highlighted with a green cross in Figure~\ref{fig:pixel_plot}, has a lower $\chi^2$ and is preferred based on the following arguments.
 Indeed, the high P solution requires high optical depths in \OI ~63\mic\ (\OI ~63\mic\ is over-predicted by a factor of up to 9, see below), while the \Avmax\ (determined as in Section~\ref{sec:CI-CO}) would be very low (< 1 mag).
 In addition, in the ionized gas we found a pressure P $= 10^5-10^6$ \cm ~K for a typical temperature of 10 000K (see Section~\ref{sec:CIIemission}).
 We find a similar pressure in the PDR, suggesting that the gas may well be in pressure equilibrium. %of the same order of magnitude for the low P solution.
However, it must also be noted that \cite{Pellegrini2011} find a pressure somewhat larger using optical lines.
Finally, a low \go\ solution results in large physical distances between the clouds and R136 (see Section~\ref{sec:3D_geo}), significantly larger (by a factor of $\sim$ 10 to 100) than found in \cite{Pellegrini2011}.

\begin{figure*}
     \centering
      \includegraphics[trim=60mm 1mm -5mm 0mm, clip, width=8cm]{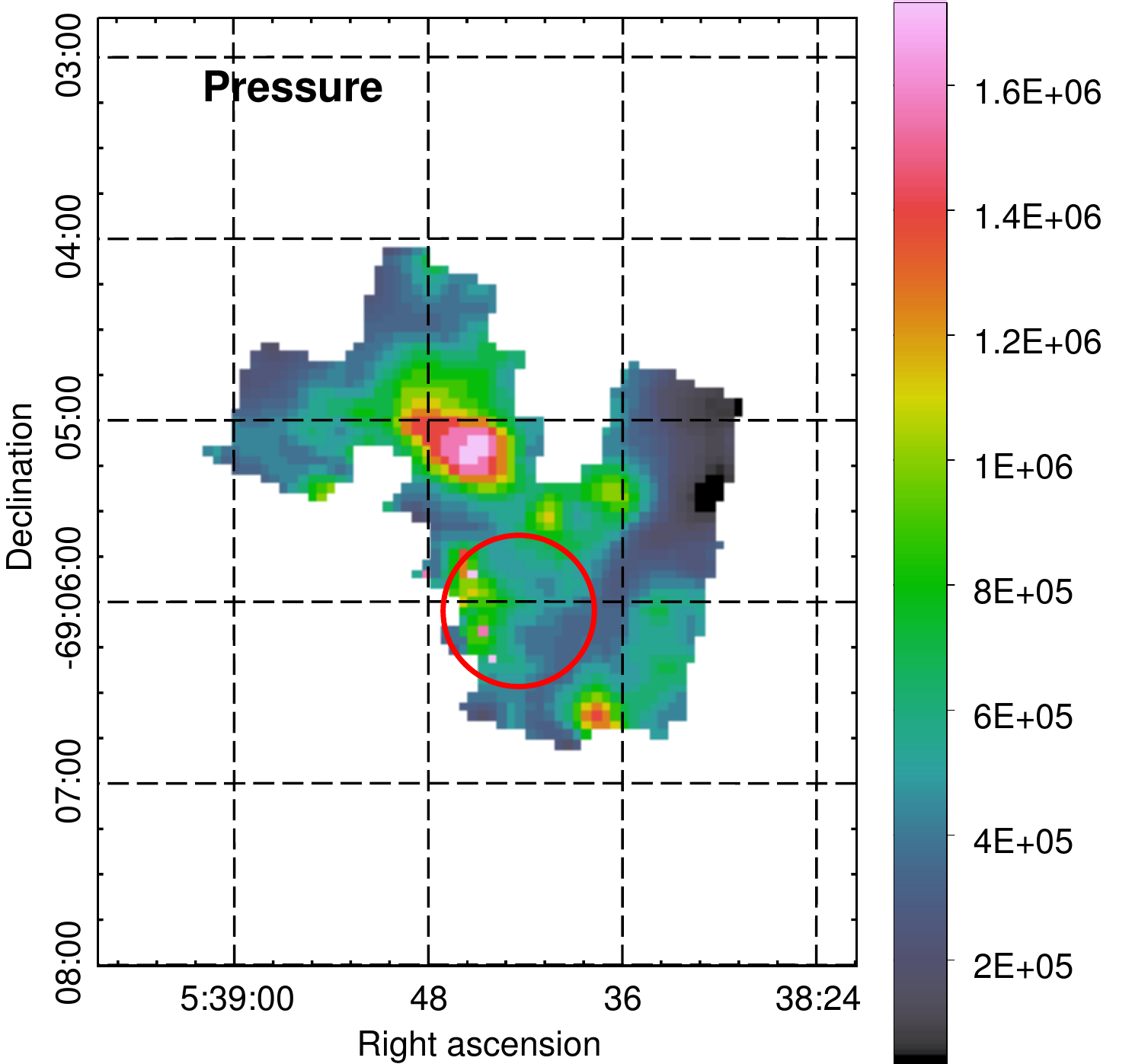}
     \includegraphics[trim=60mm 1mm -5mm 0mm, clip, width=8cm]{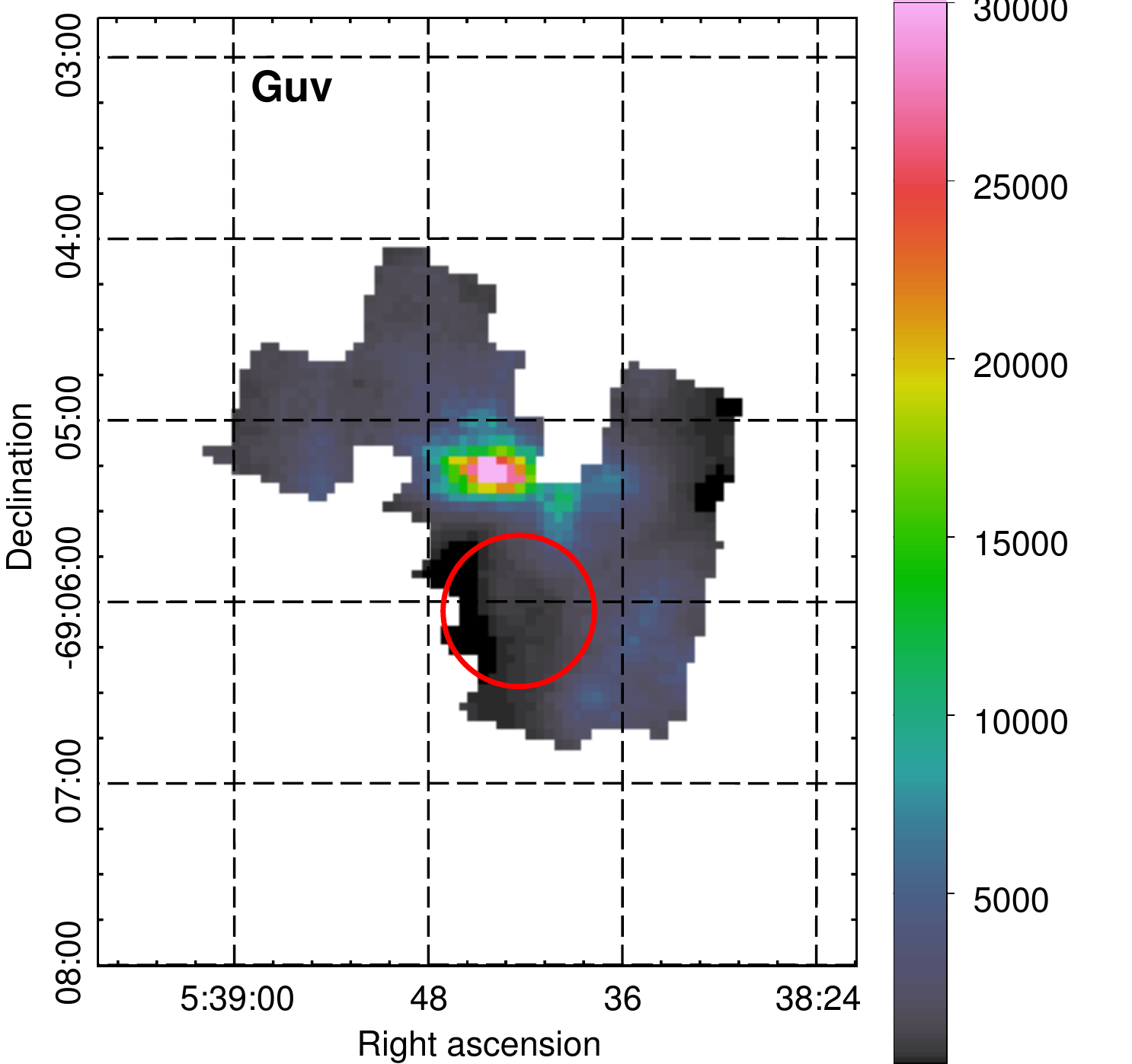}
     \caption{Left: Best pressure map (cm$^{-3}$ K) of the 30Dor region ; using (\OI ~145\mic+\CII )/\Lpdr\ and \OI ~145\mic/\CII .
      The total extinction for this model is \Avmax $=3$.
      Right: Best \go\ map (color bar in units of the Mathis field) for the same model.
      The red circle shows the location of R136.}
              \label{fig:best_Meudon05}
    \end{figure*}

We find \go\ ranging between $10^2$ and $3\times10^4$ throughout the region and P between $10^5$ \cm ~K and $1.7\times10^6$ \cm\ K, 
with $\chi^2 \sim 10^{-1}$ to $10^{-2}$ over the map\footnote{Note that this is not a reduced $\chi^2$ as we have as many constraints as free parameters.}.
The best P and \go\ maps are represented in Figure~\ref{fig:best_Meudon05}.
The peaks of \go\ and P are almost co-spatial.
The maximum is located north of R136, at the southern edge of the \OIII\ peak.
It shows that there is a void around R136 and that the radiation field is first interacting with any matter a few parsecs away from the cluster.
Details regarding the structure around R136 will be discussed in Section~\ref{sec:3D_geo}. 
The uncertainties associated with the observations, including calibration errors, lead to uncertainties on \go\ of +55\%/-40\% and on P of +/-12\%.
\newline{}

     \begin{figure*}
     \centering
     \includegraphics[trim=60mm 1mm -5mm 0mm, clip, width=8cm]{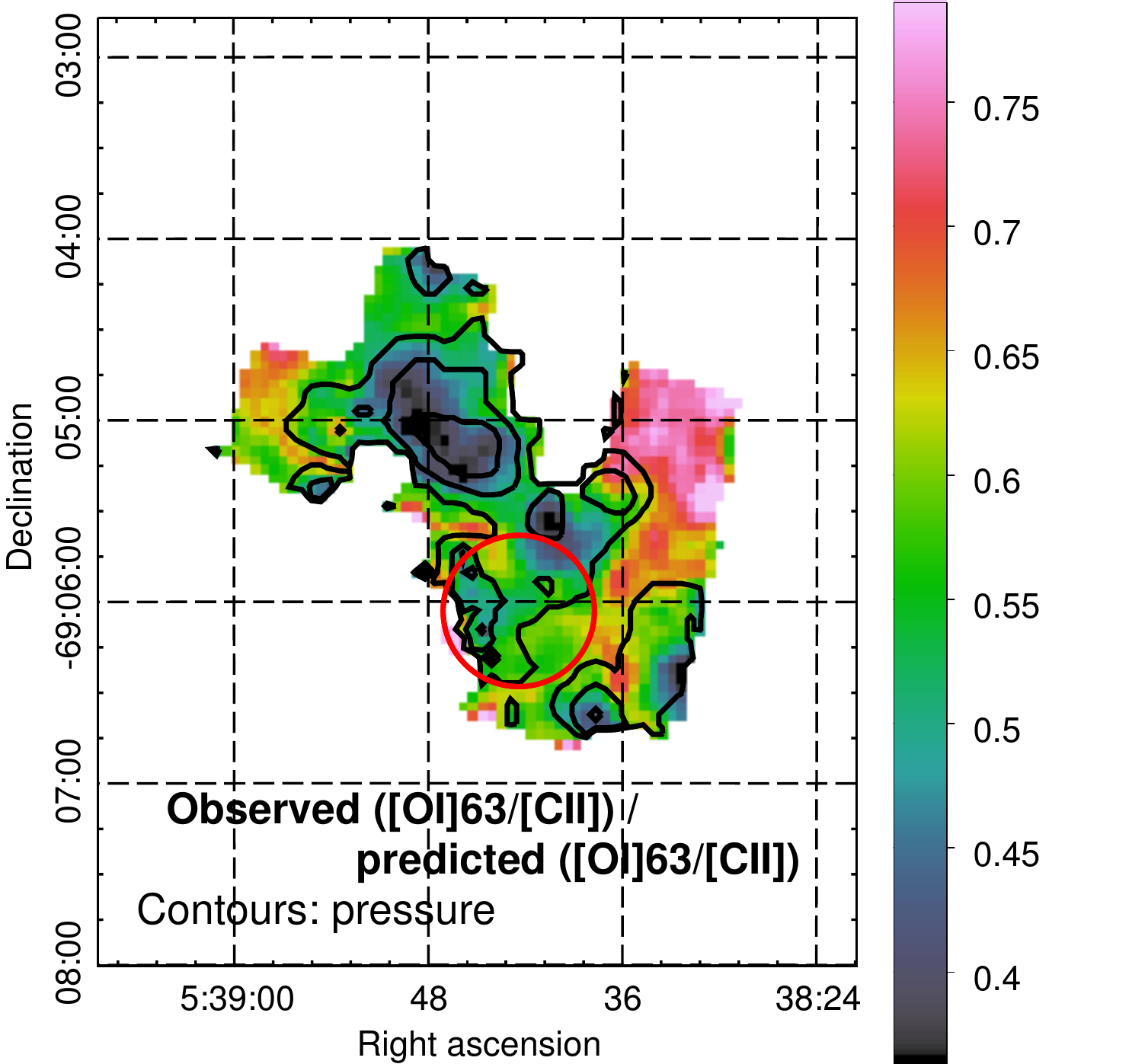}
     \includegraphics[trim= 60mm 1mm -5mm 0mm, clip, width=8cm]{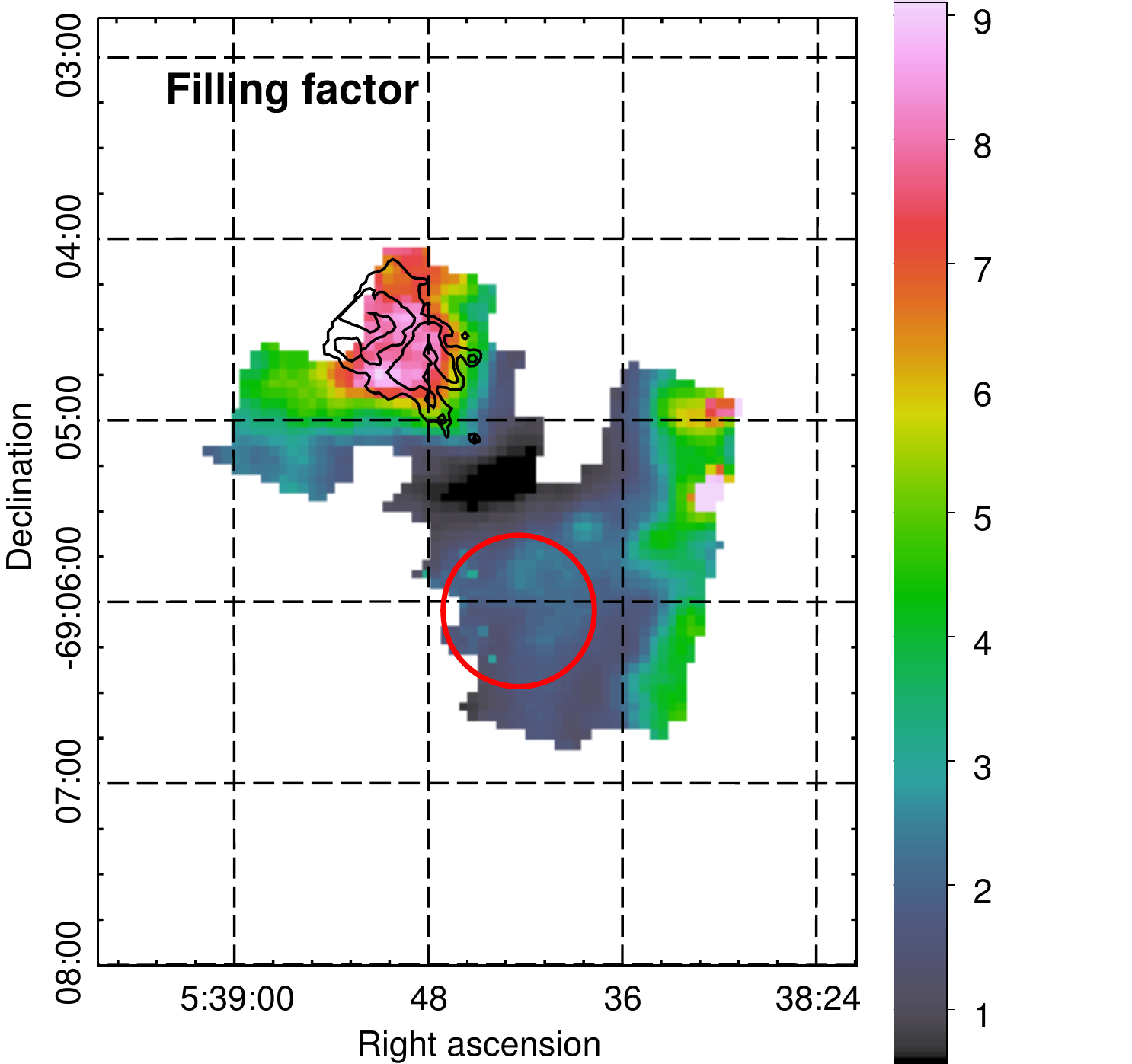}
     \caption{%Left: : Ratio of \go /P. 
     Left: Ratio of the observed \OI ~63\mic/\CII\ ratio over the predicted \OI ~63\mic/\CII\ ratio (color bar). In black are the pressure contours, between $4.5\times10^5$ and $1.3\times10^6$ cm$^{-3}$ K.
     The red circle shows the location of R136.
     Right: Filling factor $\Phi_{\rm A}$ between 0.6 and 9 (color bar). The lowest values are at the peak of \go . The black contours represent the $^{12}$CO(2-1) emission from ALMA \citep{Indebetouw2013}
     }
              \label{fig:factorOI63}
    \end{figure*}

The value of the observed ratio \OI ~63\mic/\CII\ is lower than the ratio predicted by the model based on \CII , \OI ~145\mic\ and \Lpdr\ by a factor of 1.3-2.5.
This discrepancy between the predicted and the observed values may be due to optical depth effects of the \OI ~63\mic\ line \citep{Tielens1985, Abel2007} or to absorption by cold gas along the line of sight \citep{Liseau2006}.  
As we can see from Figure~\ref{fig:factorOI63} on the left, the lowest values of the ratio $\displaystyle \frac{(\text{\OI ~63\mic}/\text{\CII})_{\text{observed}}}{(\text{\OI ~63\mic}/\text{\CII})_{\text{predicted}}}$ correspond to the locations of the highest pressure. 
Since the difference between the observations and the model prediction seems to correlate spatially with the pressure here, it is probably not due to foreground absorption, but most likely from local effects.
We can also note that the observed ratio \OI ~145\mic/\OI ~63\mic\ is larger than 0.1 in many pixels, which is an indication of optical depth effects in \OI ~63\mic\ \citep{Tielens1985}.
In the best-solution model, the predicted ratio \OI ~145\mic/\OI ~63\mic\ is 0.04.
The model accounts for the opacity of the lines for one cloud, but not between several clouds along the line of sight.
If we examine one of the best solutions for region B (\CII\ peak in Figure~\ref{fig:3colors_2}), the model predicts an opacity of 0.6 for \OI ~63\mic\ at $A_{\rm V} = 1$ while that of \CII\ is still below 0.1.
Thus, if we have several components along the line of sight, the \CII\ intensity can be multiplied by the number of components, while the \OI ~63\mic\ intensity increases less than linearly.

       \subsection{Determination of \Avmax\ using \CII, \CI\ and CO}
       \label{sec:CI-CO}
     
\begin{figure}
     \centering
      \includegraphics[trim=0mm 0mm 0mm 0mm, clip, width=9cm]{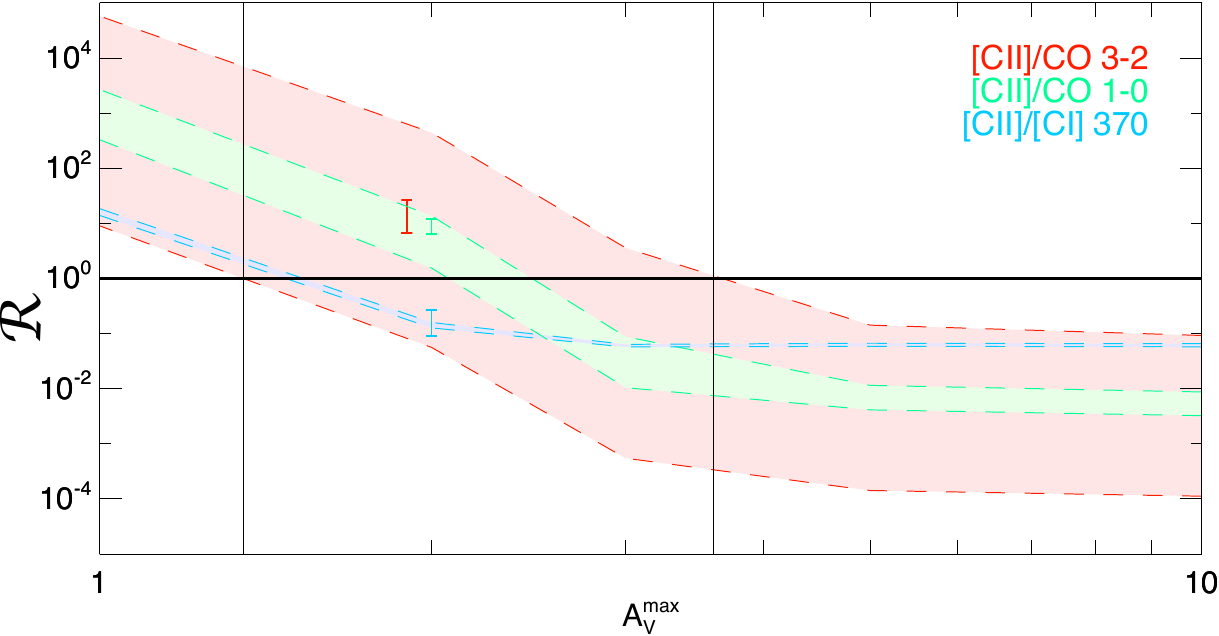}
      \caption{ Same as Figure~\ref{fig:Av_dependance}, for the ratios \CII/CO(3-2), \CII/CO(1-0) and \CII/\CI ~370\mic .
      }
              \label{fig:30dor_Av}
   \end{figure}

    In this section we investigate the influence of \Avmax\ (that is, the total depth of the cloud, in magnitude) on the line ratios predicted by the Meudon PDR code. 
    In Figure~\ref{fig:30dor_Av}, we show the ratios $\mathcal{R}$ between several modeled line ratios and their observations for simulated clouds of different total depths.
    While the ratios \OI/\CII, (\OI+\CII)/$\fir$ and \CI ~609\mic/\CI ~370\mic\ are not very sensitive to \Avmax\ (Fig.~\ref{fig:Av_dependance}) , the ratios \CII/\CI\ and \CII/CO vary by several orders of magnitude with the total depth of the simulated cloud, because they involve tracers that originate from different depths into the cloud. 
 The observed ratios \CII/\CI , \CII/CO(1-0) and \CII/CO(3-2) can be reproduced with a cloud of \Avmax $\sim 1-3$ mag per pixel of 30\arsec\ (7.2 pc) as shown in Figure~\ref{fig:30dor_Av}, with \go\ and P determined as described in Section~\ref{sec:results}.
   However we keep in mind that the determination of this parameter is strongly dependent on the geometry of the model. This will be discussed in more details in Section~\ref{sec:size}.

        \subsection{Isobaric versus isochoric case}
    \label{sec:isobar}
    
    We considered a constant density model to compare our results with our isobaric model and with previous PDR model results. %those of \cite{Poglitsch1995}.
    We use the same set of line ratios and the same $\chi^2$ method to find the best radiation field and density predicted by an isochoric model. % (see the results in appendix~\ref{app:isochore}).
    
We find a similar map for \go\ compared to our isobaric model, with a maximum of 20\% difference between both cases.
For the isochoric model, the density ranges between $3\times10^2$ to $1.4\times10^4$ cm$^{-3}$, with a spatial distribution quite similar to P in the isobaric case.
 This range of values is similar to those found by previous studies (e.g. \citealt{Poglitsch1995, Bolatto1999, Rollig2006, Pineda2012}).
From the results of the isobaric model, we determine the density at the surface of the PDR. 
This initial density is about 1/2 to 2/3 the density determined with the isochoric model.
In conclusion, the results from these two models are similar when applied to PDR line ratios.

However, the predictions of the models start to diverge deeper into the cloud, in particular for molecular lines.
Indeed, for an isobaric cloud of $A_{\rm{V}}^{\rm max}$ of 3, the density rises by about a factor of 10, as the temperature drops, between the surface of the PDR and the core of the cloud (Fig.~\ref{fig:abundance_profiles}).
This implies that the CO lines are emitted at lower $A_{\rm{V}}$ in a isobaric model compared to an isochoric model.
As a consequence, the $A_{\rm{V}}^{\rm max}$ probed by the observed CO lines is lower for an isobaric model than for an isochoric model.
For example, the $A_{\rm{V}}^{\rm max}$ probed with the low-J CO transitions (J=3--2 and J=1--0) is slightly higher for the isochoric model ($\sim 2-4$ mag) than the $1-3$ mag we find with the isobaric model.
In conclusion, choosing the isobaric case is important to reproduce the higher J transitions of CO and quantify the CO-dark gas (see paper II), but it has little consequences for the results derived in this paper.

              \subsection{Model predictions for \HH\ lines}

\HH\ is barely detected in the \spit /IRS low-resolution observations of 30Dor \citep{Indebetouw2009}.
The model-predicted \HH\ (0,0) $S$(2) emission at 12.3\mic\ is 10 to 50 times lower than the upper limit from the observations.
However, the emission from the high-resolution spectrum (Sect.~\ref{sec:spitzer}) is in better agreement (by a factor of 2 to 5) with the model prediction.

\cite{Pak1998} measure \HH\ (1, 0) $S$(1) and (2, 1) $S$(1) close to the CO (1-0) peak of 30Dor, with a beam of $81^{\prime \prime}$. 
We compare these observations to the predictions of our model at 81\arcsec\ resolution.
Our map does not fully cover the 81\arcsec\ beam.
 However, we calculate lower limits of $6\times10^{-6}$ and $3\times10^{-6}$  \escmsr\ on the \HH\ (1, 0) $S$(1) and (2, 1) $S$(1) emission respectively.
There is a good agreement with the values measured by \cite{Pak1998} ($10.8\times10^{-6}$ and $4.0\times10^{-6}$  \escmsr\ respectively) considering we are missing some fraction of their beam.
\cite{Rubio1998} present \HH\ (1, 0) $S$(1) observations at 1.16\arsec\ resolution. 
Their peak value of $4-6.10^{-5}$ \escmsr\ is also consistent with our model results.

\cite{Yeh2015} have presented an \HH\ (1, 0) $S$(1) map of the entire 30Dor nebula at 1\arsec\ resolution.
The peak \HH\ surface brightness in their region A is $9.15\times10^{-5}$ \escmsr .
Our model predicts a maximum \HH\ surface brightness of $5.3\times10^{-5}$ \escmsr\ located near the \CII\ peak, for a 12\arsec\ resolution.
This is about a factor of 2 lower than the observed value, but it may be explained by the lower resolution used in our study and the fact that the \HH\ emission could originate from clouds smaller than the PACS beam size.

%__________________________________________________________________

\section{Discussion}
  \label{discussion}

  \subsection{Filling factor}  
  \label{sec:fill_factor}

The model assumes that the PDR filling factor is unity, i.e., that the surface area of PDRs is equal to the beam area.
If only part of the beam is covered or, on the contrary, if several clouds are present along the line of sight, the radiation emitted in the \Lfir\ and in cooling lines will be different than the model prediction for optically thin lines. 
The filling factor will be, respectively, lower or higher than 1.

\go\ and P have been determined only using ratios of lines coming from the same phase of the ISM. 
Thus, they do not depend on the filling factor because each line is affected by the same factor. 
The area filling factor, $\Phi_{\rm A}$, can then be estimated from the ratio of the observed intensity over the predicted intensity (for which $\Phi_{\rm A}=1$) for an individual line.

\Lpdr\ and \CII\ are both emitted by PDRs and scale with the filling factor in the same way.
Thus, we can determine $\Phi_{\rm A}$ either with $\displaystyle \frac{\text{\CII}_{\text{observed}}} {\text{\CII}_{\text{predicted}}}$ 
 (as in \citealt{Wolfire1990}) or similarly with $\displaystyle \frac{\text{\Lpdr}_{\text{observed}}} {\text{\Lpdr}_{\text{predicted}}}$. 
This result is presented on the right panel of Figure~\ref{fig:factorOI63}.
$\Phi_{\rm A}$ is the smallest (about 0.6) near the peak of \go , in region C (Fig.~\ref{fig:3colors_2}). 
The maximum value is $\sim 9$ on the north-east of the cluster.
As stated earlier, an area filling factor greater than one means that several clouds covering the entire pixel are along the line of sight.
Note that this approach is marginally coherent near the peak of \CII\ where the maximum filling factor is $\sim 9$. 
As noted in Section~\ref{sec:results}, considering that the opacity of \CII\ is close to 0.1 at the \CII\ peak, 9 clouds on the line of sight will add up to an opacity close to 1.
As a consequence, the \CII\ intensity will not add up linearly and this may call into question our assumption of optically thin \CII\ at the \CII\ peak location.
Different PDR viewing angles can also affect this scenario and the number of components we derive is an upper limit.

This spatial distribution of the filling factor reflects the idea that the PDR clouds are smaller and/or less numerous where the radiation field is higher. 
The region mapped with ALMA by \cite{Indebetouw2013} corresponds to the region where the filling factor of PDRs is the largest in the map (see Figure~\ref{fig:factorOI63}).
The CO emission is highly structured, showing dense ($10^3$--$10^5$ cm$^{-3}$) clumps and filaments, on sub-parsecs scales, covering approximately 10\% of the area of the map. 
They observed significant photodissociation of CO by radiation penetrating between the dense clumps, which can explain the increased \Xco\ factor determined for unresolved low metallicity clouds.
\citeauthor{Indebetouw2013} measure an \Xco\ factor two times the solar-metallicity value in the dense CO clumps.
This value can be even higher when possible interclump \HH\ is included, for unresolved studies.
\newline{}

 \subsection{Clouds geometry} 
  \label{sec:size}

     \cite{Indebetouw2013} observed the giant molecular cloud 30Dor-10, north of R136 with ALMA during Cycle 0 (it includes our regions B and C from Figure~\ref{fig:3colors_2}).
   In the region mapped with ALMA, they measure a maximum diameter of 1.2~pc for the CO(2-1) clumps and a filling factor of about 10\%.
  This could mean that the CO luminosity is dominated by small, bright clouds in the CO beam and the medium could be characterized by clumps/filaments of low volume filling factor, with a physical size much smaller than our pixel size, implying that a plane parallel geometry together with the high filling factor determined with \CII\ may not  be adapted to our observations. %which is in contradiction with the plane parallel geometry of the model and the high filling factor determined with \CII .
  
Indeed, with the \Avmax , \go\ and P previously determined in Sections~\ref{sec:CI-CO} and \ref{sec:results}, we find that several plane parallel clouds are required to reproduce the observed absolute intensities.
Each of these clouds have a physical size between 0.2 and 3 pc, and an internal CO layer (estimated where the CO abundance is at least 50\% of the maximum CO abundance) that ranges between 0.06 to 0.5 pc. 
The derived size scale is compatible with the ALMA observations.
  These results however depend on two main assumptions: the plane parallel geometry and the uniformity of PDR clouds in each pixel.
  
  With a plane geometry, the \Avmax\ of the simulated clouds determined here by comparison of the \CII\ and CO emissions can be considered as a lower limit. % of the actual cloud. %due to the plane parallel geometry of the model.
  Indeed, adopting a different geometry by post-processing the results of the model and wrapping the plane-parallel results on a spherical geometry will result in a larger \Avmax .
  More details about this method can be found in Appendix~\ref{app:wrap}.
  When assuming a spherical geometry, bathed in an isotropic radiation field, a higher \Av\ ($\sim20$ mag at the center of the cloud) is needed to reproduce the CO observations.
 Such extinction corresponds to spherical clouds with diameters about four times larger than the size of the optimal plane parallel clouds.
  Nevertheless the results derived in Section~\ref{sec:results} for \go\ and P are independent on the assumed geometry and remain unchanged.
  
\begin{figure}
     \centering
      \includegraphics[trim= 15mm 1mm 15mm 9mm, clip, width=8cm]{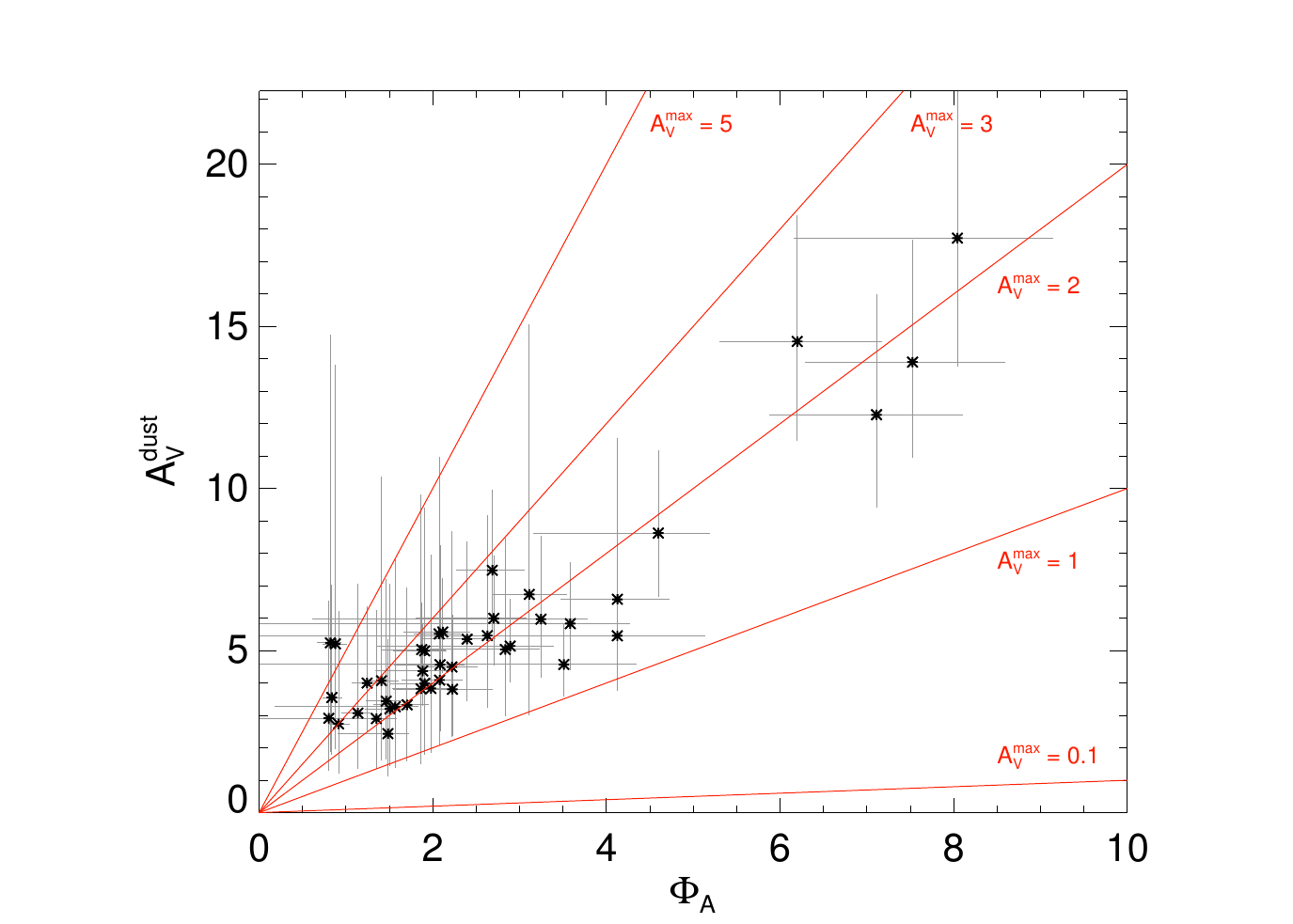}
       \caption{The extinction derived from the dust mass \Avdust\ is plotted against the filling factor \PhiA . 
                       The diagonal lines show different values of the extinction in individual PDR clouds (\Avmax ). 
                       The data points agree with \Avdust\ for \Avmax $\sim1-3$.}
              \label{fig:Avdust}
    \end{figure}

  We note that the extinction parameter can be determined independently from the dust mass surface density, at a resolution of 22" (Sect.~\ref{sec:IRmaps}). 
We can then compare \Avdust\ derived from the dust map to \Avmax\ derived using the \CII /\CI\ and \CII /CO ratios in our PDR models (Sect.~\ref{sec:CI-CO}). 
On the one hand, \Avdust\ corresponds to the integrated extinction along the line of sight in each pixel. 
On the other hand \Avmax\ is calculated for individual clouds, so for a meaningful comparison we need to scale by the number of components (i.e., the filling factor \PhiA ; Sect.~\ref{sec:fill_factor}). 
While \PhiA\ is calculated at a relatively high spatial resolution (12\arcsec ), the \Avmax\ for individual clouds is not well constrained due to the poor 
spatial resolution (42\arcsec) and coverage of the \CI\ and CO maps. 
In Figure~\ref{fig:Avdust}, we plot \Avdust\ as a function of \PhiA\ (recalculated at a resolution of 22\arcsec), with several tracks showing the scaling with different \Avmax\ values. 
There is a remarkable correlation between \Avdust\  and \PhiA , with most data points lying between the \Avmax\ = 1 and 3 lines, in agreement with the values determined with a plane-parallel geometry (Sect.~\ref{sec:CI-CO}). 
We emphasize that the determinations of  \Avmax\ and \PhiA\ are quite independent from the determination of \Avdust , 
as the two quantities are derived from different physical processes, and constrained by different observational sets. 
From Figure~\ref{fig:Avdust}, we infer that most clouds share similar \Avmax\ values at a spatial resolution of 22\arcsec. 
Moreover, the good agreement between \Avdust\ (total extinction) and \Avmax\ (derived in the neutral gas of PDRs) suggests that 
(1) there is no significant contribution from the ionized gas in the \Avdust\  determination, 
and (2) there is no significant contribution from foreground/background gas not associated to 30Dor and not accounted for by our PDR models. 
Our result not only strengthens the \Avmax\ range obtained in Section~\ref{sec:CI-CO}, but also suggests that this range remains valid at a resolution of 22\arcsec. 
While this result seems to favor the plane-parallel geometry over the spherical geometry, we wish to stress that we cannot easily derive the effective extinction corresponding to spherical clouds observed in any given pixel and that the large extinction probed at the center of the spherical cloud is not representative of this effective extinction.

Finally, we discuss the possible existence of clumps embedded in an 
interclump medium.
 Since the filling factor determined with \CII\ in Section~\ref{sec:fill_factor} may characterize an interclump medium surrounding small CO clumps, it is possible that the \CII\ and the CO emission are not associated with the same structures.
  For example, a recent study of the N159 region by \cite{Okada2015} find that up to 50\% of \CII\ cannot be associated with the CO emission based on the velocity profiles of the lines.
  Testing the clumpiness of the medium would require to adopt a distribution of clouds of different sizes and pressures, illuminated by a central source and including the effect of scattering and shielding, for each pixel of the map.
  Unfortunately, such a scenario cannot be properly modeled yet due to the lack of observational constraints.

   \subsection{3D distribution of the gas}
    \label{sec:3D_geo}
 
\begin{figure}
     \centering
      \includegraphics[trim= 2mm 0mm 2mm 2mm, clip, width=8cm]{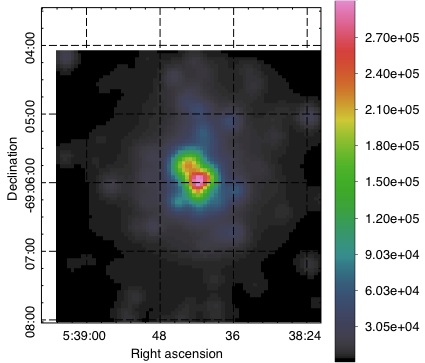}
       \caption{Intensity of the emitted radiation field \gstar\ in the plane of R136, at 12\arsec\ resolution (color bar in units of the Mathis field).}
              \label{fig:gstar}
    \end{figure}

 In this section we determine the physical distance between the clouds and the ionizing sources to reconstruct the 3D distribution of the gas. 
 This is done by comparing the incident \go , predicted by the Meudon PDR code, and the emitted radiation field, \gstar .
We define \gstar\ as the FUV radiation field computed from the known massive stellar population from the literature.
 We use catalogs of stars from \cite{Crowther1998} and \cite{Selman1999}, including O and B stars and Wolf-Rayet (WF) stars.
 We use the temperature of the brightest optical sources and integrate over a blackbody 
 between 912 and 2400 $\AA$ to be consistent with the definition of \go\ in the Meudon PDR code.
 We then use a $\displaystyle \frac{1}{R^2}$ relation, where $R$ is defined as the physical distance from the center of the cluster, to calculate an average \gstar\ in each pixel of a cube centered on R136.
 We first make the assumption that all of the stars lay on the same plane, and derive the \gstar\ presented in Figure~\ref{fig:gstar}.
 \gstar\ is then the maximum incident radiation field we expect on each point of the cube since no absorption is taken into account.
 
 The ratios used to constrain \go\ and P are independent of the filling factor, since \OI ~145\mic , \CII\ and \Lpdr\ are supposed to be almost co-spatial in the PDR model.
Thus, \go\ determined with the model is also independent of the filling factor. 
In that case, the incident radiation field, \go , should be equal to the emitted radiation field \gstar, modulated by the distance to the R136 plane, assuming there is no absorption between the ionizing stars and the PDR (i.e. the distance to the plane calculated here is an upper limit on the actual distance).
 We can determine simultaneously four parameters : \go, P, $\Phi_{\rm A}$ and $z$ (the distance to the plane of R136)
for each pixel of the map, using the following relations : 
\begin{align}
\text{\Lpdr}_{obs} &= \Phi_{\rm A} \times \text{\Lpdr}_{pred}\\
{\rm G}_{stars} &= \text{\go} \times \frac{z^2+d^2}{d^2}\\
I_{j} &= M_{j}
\end{align}

where $d$ is the projected distance between a pixel and the central source, $\sqrt{z^2 + d^2} = R$.
$I_{j}$ are the observed values of the line intensity ratios $j$ and $M_{j}$ the ratios calculated by the PDR model (in \escmsr ).
\go\ and \gstar\ are in units of the Mathis field.
\go\ and P are calculated as described in Section~\ref{sec:results}, using Equation 4.

\begin{figure}
     \centering
      \includegraphics[trim= 60mm 0mm 10mm 0mm, clip, width=8cm]{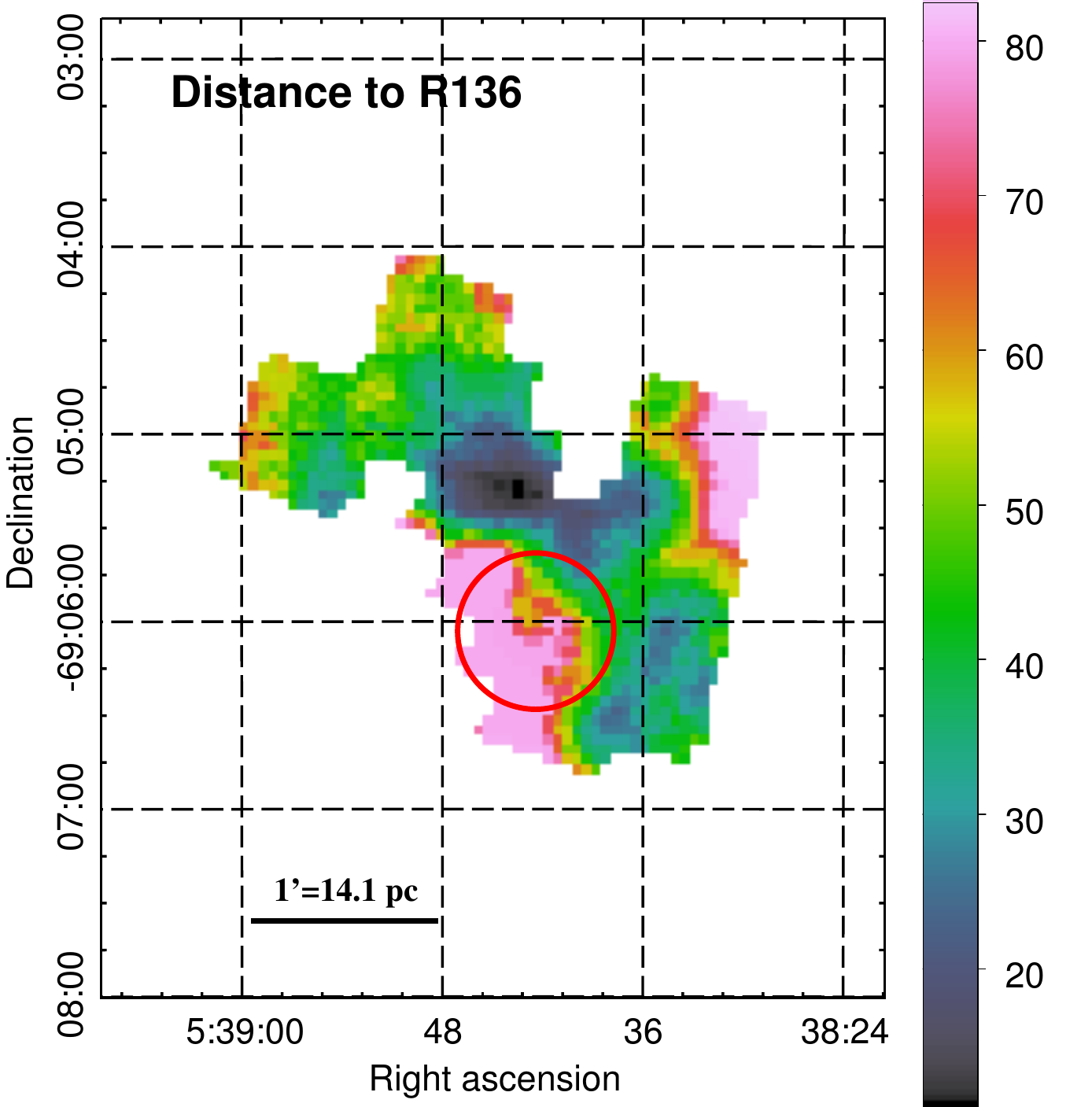}
     \caption{Physical distance $R$ (in pc) of the surface of the PDR  to the center of R136.
                     The red circle shows the location of R136.}
              \label{fig:distR136}
    \end{figure}

The physical distance $R$ between PDR clouds and R136 is presented in Figure~\ref{fig:distR136}.
 It ranges from $\sim$ 11 to 80 pc.
 The gas located close to R136 in the projected view is actually 40 to 80 pc away from the cluster.
 The bright arm-like structure in \OIII , which, when projected, appears further from R136, is much closer to the star cluster as it is almost on the same plane.
 This is consistent with the distance calculated in \cite{Pellegrini2011} for the ionized gas using optical observations.

The physical distance described above was derived by assuming that all of the stars are located in the same plane. 
For comparison, we have also considered a random distribution of the stars in the perpendicular direction, but maintaining a high density of stars within a 6 pc radius around the center of R136 in order to reproduce a spherical distribution. 
We performed a Monte-Carlo simulation and calculated distances similar to our previous determination (< 40\% difference throughout the map). 
This is not surprising as most stars in R136 are in fact located in a $\sim$ 6 pc radius sphere. 
We used the Monte-Carlo simulation to estimate a typical uncertainty on the physical distance of $\sim$ 4 pc.

    \subsection{Porosity of the ISM}
    \label{sec:OIIICII_FIR}
 %%%%%%%      OIII/CII vs FIR

\begin{figure}
      \centering
       \includegraphics[trim = 5mm 1mm 0mm 2mm, clip, width=8.7cm]{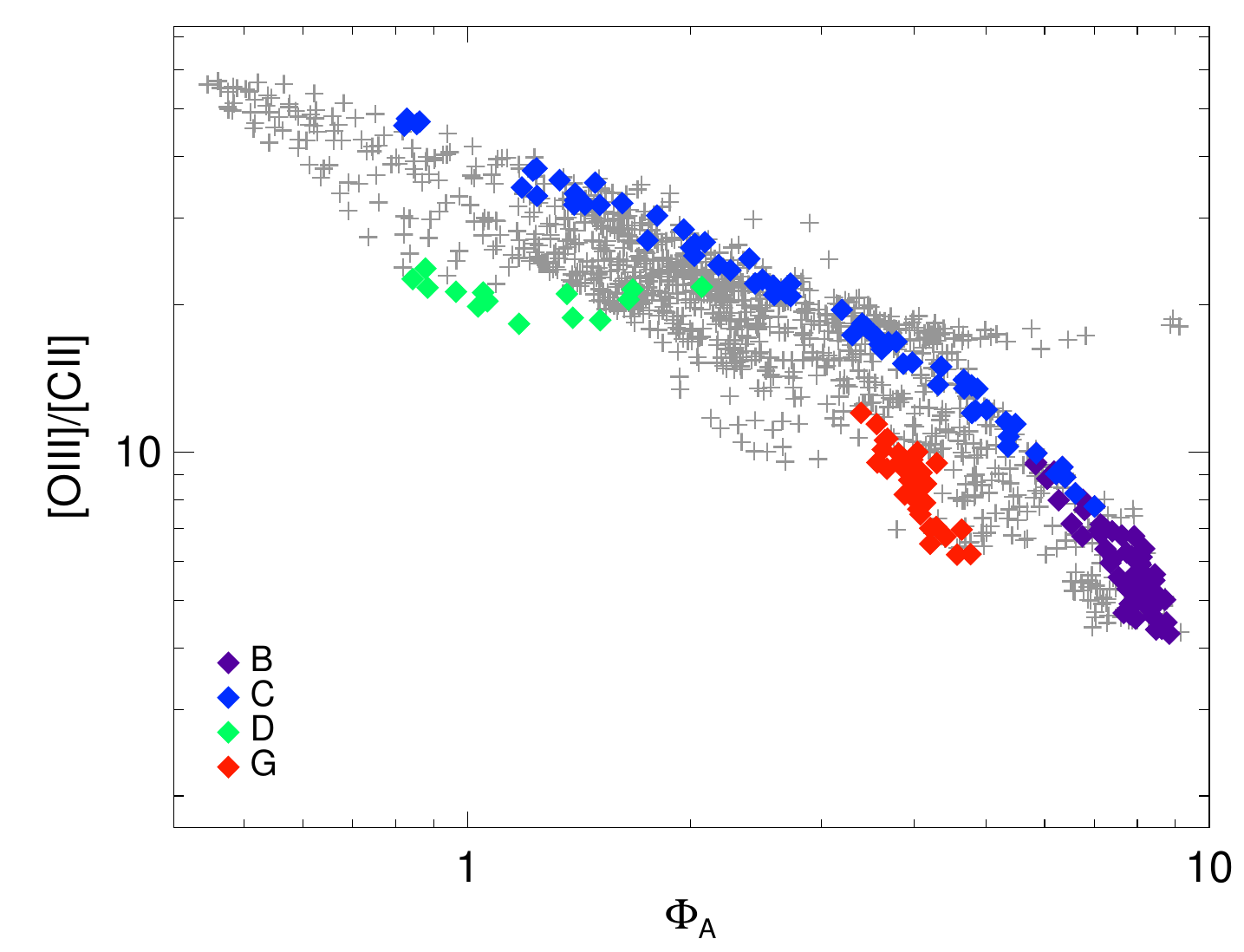}
      \caption{
                     Correlation between \OIII/\CII\ and the PDR area filling factor. 
                     The color symbols are associated with the regions defined in Figure~\ref{fig:3colors_2}. 
                     The grey symbols are all of the other pixels of our 30Dor map.
                     } 
      \label{fig:oiiicii_Lfir}
\end{figure}

The \OIII ~88\mic\ line emission is detected over large spatial scales in 30Dor, as it had been already noticed in other extended sources (e.g. in N11 by \citealt{Lebouteiller2012})
and, together with the small scale CO clumps we see, may be indicative of a highly porous region.
There is probably mixing of the ionized and neutral phases of the ISM throughout all spatial scales.
This idea is supported by Figure~\ref{fig:oiiicii_Lfir}, which 
shows the ratio \OIII/\CII\ as a function of $\Phi_{\rm A}$.
We can see that these two quantities are strongly correlated.
Indeed, a low area filling factor $\Phi_{\rm A}$ in a given pixel implies a small volume fraction occupied by PDRs. 
This means that there is less matter per unit volume to absorb the UV radiation, and this radiation is able to travel further away.
Region B from Figure~\ref{fig:3colors_2} (violet), at $\sim 40$ pc from R136, has the lowest \OIII /\CII\ ratio and the highest PDR filling factor.
This is also where the \CII\ peaks and the CO clumps reside (see contours in Figure~\ref{fig:factorOI63}).
In contrast, the ratio \OIII /\CII\ is high (> 10) in region D (green), which is one of the furthest region from R136, with a low $\Phi_{\rm A}$.
The blue points of region C, which is the \OIII\ peak and near the \go\ and P peaks, show a large range of \OIII /\CII\ and $\Phi_{\rm A}$.
This is a region of widely varying ISM conditions.
Note that the ratio \OIII/\CII\ as a function of the physical distance R 
or \go\ is a scatter plot, highlighting the fact that the proximity of the ionizing source is not the only controlling factor of the structure of the ISM.

Over the mapped area of 42 pc $\times$ 56 pc the PDR filling factor and the \OIII /\CII\ ratio vary over one order of magnitude.
The decrease in dust abundance and intense UV photons from the SSC conspire together to shape the surrounding porous ISM, filling it with hard photons and relatively small filling factor of PDR clumps.

\subsection{Other sources of excitation}
\label{sec:excitation}

\begin{figure}
     \centering
      \includegraphics[width=8cm]{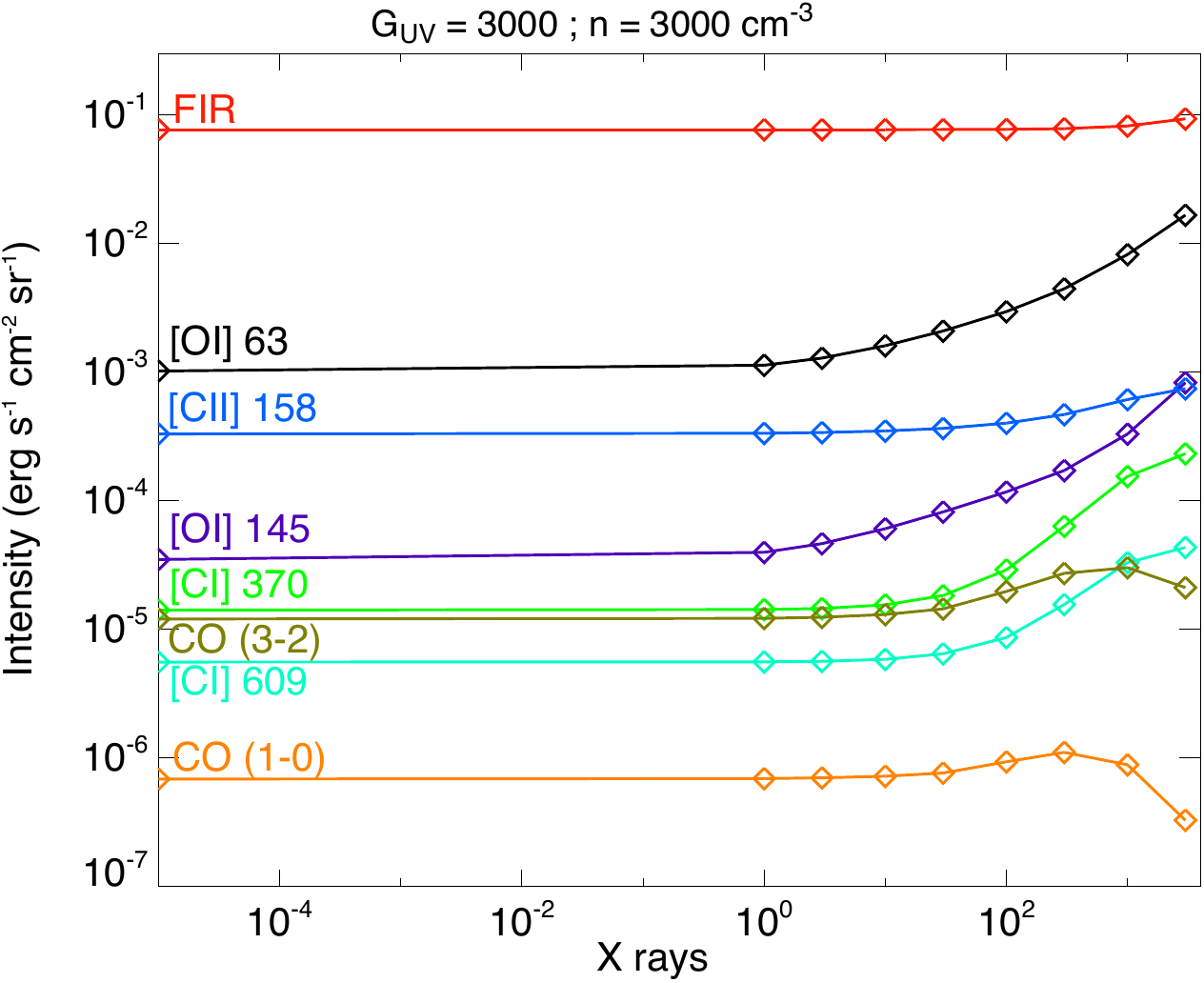}
     \caption{Influence of the X-ray luminosity on the intensities of individual lines for a typical isochoric model with \go\ $= 3\times10^3$ and n~$=3\times10^3$~cm$^{-3}$. 
                     \gx\ is the X-ray luminosity integrated between 0.2 and 8 keV, given in unit of the Habing field \citep{Habing1968}.
                     This is to be compared with our results for \go\ (see Section~\ref{sec:excitation}). }
              \label{fig:xrays}
    \end{figure}

No evidence of any shock tracers has been found in 30Dor for now by previous studies \citep{Indebetouw2009, Yeh2015}.
We investigate here the possibility for X-rays to be an important source of excitation of the gas in 30Dor.
\cite{Townsley2006a} have studied the population of X-ray point sources (energy between 0.5 and 8 keV) in a $17^{\prime} \times 17^{\prime}$ field around R136 with \chandra .
Spectral fitting is performed on the brightest sources (49 in total).
In particular, they determine a total X-ray luminosity, corrected for the absorption, of $10^{36.95}$ erg s$^{-1}$ for the brightest source in R136 (Mk 34), 
with an X-ray flux of $2\times10^{-3}$ \escmsr\ at a distance of 20 pc away, approximately at the ionization front. 
As presented in Figure~\ref{fig:xrays}, such a low \gx , compared to the values of \go , does not have an important effect on the intensity of the individual lines for a given \go\ and density.
\newline{}

  \subsection{If 30Dor were unresolved}
  
Studying resolved nearby galaxies can help us understand more distant unresolved targets.
  At a distance of 900 kpc, the entire region of 30Dor mapped in \OI ~145\mic\ with PACS (covering 56 pc $\times$ 70 pc) would fall in only one spaxel of the PACS spectrometer.
  This distance is comparable to the distance of the Andromeda galaxy ($\sim$780 kpc).
  
  We integrate all of the tracers we have and perform the same study at this lower resolution, with only one pixel.
  Using the same technique as Section~\ref{sec:results}, we find that all of the \CII\ emission originates from the PDRs.
  Using the Meudon PDR code, we determine \go\ $ = 1390$, P $ =6.18\times10^5$ cm$^{-3}$ K and $A_{\rm{V}} = 1-2$.
  This is representative of a region of moderate \go\ and P in our detailed spatial study such as region G in Figure~\ref{fig:3colors_2} or the ISM north of region B.
 
  Even though our map is relatively small and centered on R136, the global solution is already biased toward the solution corresponding to regions with relatively low P and low \go .
  However, we have to keep in mind that the regions targeted with PACS are the most luminous around R136, especially for the \OI ~145\mic\ line.
 This implies that if we were to integrate an even larger area, the result would be presumably biased even more to this diffuse and low \go\ regime.
  These two parameters could be even lower if we include more diffuse regions, which is the subject of a subsequent study in the LMC.

%______________________________________________________________

\section{Conclusions}
 \label{Conclusion}

We have studied the ISM properties in the extreme environment of 30Dor in the LMC, around the super star cluster R136.
We summarize our results as follow:

   \begin{enumerate}
      \item We have presented and analyzed new \her /PACS observations of 30Dor in \CII ~158\mic , \OI ~63 and 145\mic , \NII ~122 and 205\mic\ and \OIII ~88\mic\ over a 56$\times$70 pc region of 30Dor.
                All of these lines are well detected and provide diagnostics on the structure of the ISM.
                
     The \OIII\ line is the brightest of the FIR lines, ranging from 2 to 60 times more luminous than \CII .  
     We propose that the \OI ~63\mic\ line is optically thick
     throughout the mapped region.
     We find that the \CII /CO(1-0) luminosity ranges between $5\times10^4$ and $3\times10^5$ throughout the map. 
     This range is larger than the broad range of ratios found from integrated DGS dwarf galaxies of \cite{Cormier2014}. 
    
    \item Using the \spit\ and \her\ MIR to submm photometric observations, we model the full dust SED spatially around 30Dor and derive the infrared luminosity map. 
    We find that the \CII\ intensity ranges from 0.1 to 1\% of the observed \Lfir\ throughout the region mapped. 
                
      \item Based on the electron density of the ionized gas (10 to 100 cm$^{-3}$) and on the high value of the \CII/\NII\ ratio, we have determined that at least 90\% of the \CII\ emission originates from the PDRs, which makes the \CII\ intensity a valuable constraint for our PDR modeling.

     \item We have decomposed the \Lfir\ map to separate the component associated with the ionized gas from the PDR-only component. 
      We find, in places, that $\sim$ 70\%  of the \Lfir\ is not from PDRs, but associated with the ionized gas component, which we remove for the PDR modeling. We emphasize caution when applying the total \Lfir\ to PDR models without considering the origin of the \Lfir .
                
      \item From the ratios (\OI ~145\mic +\CII )/\Lpdr\  and \OI ~145\mic/\CII\ we have determined the spatial distribution of the radiation field and the pressure with the Meudon PDR code.
                \go\ ranges between $10^2$ and $3 \times 10^4$ (in units of the standard radiation field defined in \citealt{Mathis1983}) and P ranges between $10^5$ and $1.7\times10^6$ cm$^{-3}$ K.

      \item The total depth of the clouds is determined by including the ratios \CII/\CI\ or \CII/CO in the modeling assuming that all of these tracers are associated with the same structures.
               We showed that in the 30Dor region, \Avmax\ in 30\arcsec\ pixels is $\sim$1-3 mag.
               This value should be considered as a minimum value due to our assumption of  a plane-parallel geometry.

      \item We have conducted a 3D model of the PDR gas around R136.
      Comparison of the incident radiation field determined from our PDR model, \go , with the emitted radiation field, \gstar, reveals that the PDR gas is distributed at various distances ranging between 20 to 80 pc from the excitation source, R136.
      
            \item The PDR area filling factor ranges between 0.6 (at the peak of \go ) and 9 (at the peak of \CII ).
            The high value of the \OIII/\CII\ ratio and its tight correlation with the filling factor rather than with the distance, highlight the porosity of the ISM, filled with hard photons around relatively small PDR clumps.
            
   \end{enumerate}
The combined effects of a half-solar metallicity gas with the intense excitation source R136 create the extreme environment we see in 30Dor.
It has been shown that the structure of the gas in this region is dominated by photoionization.
X-rays or shocks are not needed to reproduce the observed line intensities.
Based on our findings in the present study, we speculate that the small size of the CO core inside the PDR clouds could explain the high \CII/CO ratio we observe in low metallicity environments.
The high value of the \OIII\ emission line suggests that a highly porous medium is a characteristic of the gas in low-metallicity dwarf galaxies.

\begin{acknowledgements}
      The authors would like to thank Annie Hughes for providing the CO(1-0) data and Akiko Kawamura for the CO(3-2) data. 
      This research was made possible through the financial support of the Agence Nationale de la Recherche (ANR) through the programme SYMPATICO (Program Blanc Projet (NR-11-BS56-0023) and also through the EU FP7.
PACS has been developed by MPE (Germany); UVIE (Austria); KU Leuven, CSL, IMEC (Belgium); CEA, LAM (France); 
      MPIA (Germany); INAFIFSI/OAA/OAP/OAT, LENS, SISSA (Italy); IAC (Spain). 
      This development has
been supported by BMVIT (Austria), ESA-PRODEX (Belgium), CEA/CNES
(France), DLR (Germany), ASI/INAF (Italy), and CICYT/MCYT (Spain).
SPIRE has been developed by Cardiff University (UK); Univ. Lethbridge
(Canada); NAOC (China); CEA, LAM (France); IFSI, Univ. Padua (Italy);
IAC (Spain); SNSB (Sweden); Imperial College London, RAL, UCL-MSSL,
UKATC, Univ. Sussex (UK) and Caltech, JPL, NHSC, Univ. Colorado (USA).
      This development has been supported by CSA (Canada); NAOC (China); CEA,
CNES, CNRS (France); ASI (Italy); MCINN (Spain); Stockholm Observatory
(Sweden); STFC (UK); and NASA (USA).
      SPIRE has been developed by a consortium of institutes led by Cardiff
Univ. (UK) and including: Univ. Lethbridge (Canada); NAOC (China); CEA,
LAM (France); IFSI, Univ. Padua (Italy); IAC (Spain); Stockholm Observatory
(Sweden); Imperial College London, RAL, UCL-MSSL, UKATC, Univ. Sussex
(UK); and Caltech, JPL, NHSC, Univ. Colorado (USA). 
      This development has
been supported by national funding agencies: CSA (Canada); NAOC (China);
CEA, CNES, CNRS (France); ASI (Italy); MCINN (Spain); SNSB (Sweden);
STFC, UKSA (UK); and NASA (USA).
\end{acknowledgements}

%__________________________________________________________________

\appendix

\section{PACS observations}
\label{app:pacs}

\begin{landscape}
\small{
\begin{table}
\begin{center}
\begin{tabular}{| c | c | l | c | c | c | c | c | c |}
     \hline
     OBSID            &  Coordinates                             &  Lines                         &   Observation date  &  Exposition time (s)  &  Rasters nb &  Mode\\ %&  repetitions  &  cycles 
     \hline
    1342222085  & 5h38m35,00s  -69d05m39,0s  &\OIII $_{88}$,  \CII $_{158}$    & 740                              & 2513,0                                &  4  &  faint line\\ %&  1  &  1  
     \hline
    1342222086  & 5h38m48,00s  -69d06m37,0s  &\OI $_{63}$                        & 740                             & 908,0                                   &  2  &  faint line\\   %&  1  &  1
     \hline
    1342222087  & 5h38m58,00s  -69d04m43,0s  &\OI $_{63}$                        & 740                             & 1322,0                                 &  4  &  faint line\\   %&  1  &  1
     \hline
     1342222088  & 5h38m58,00s  -69d04m43,0s &\OIII $_{88}$,  \CII $_{158}$    & 740                             & 2515,0                                &  4  &  faint line\\   %&  1  &  1
     \hline
     1342222089  & 5h38m48,00s  -69d06m37,0s &\OIII $_{88}$,  \CII $_{158}$    & 740                             & 1703,0                                &  2  &  faint line\\     %&  1  &  1     
     \hline  
     1342222090  & 5h38m56,66s  -69d04m56,9s &\OI $_{145}$                        & 740                             & 451,0                                  &  1  &  faint line, pointed\\ % &1&1     
     \hline
     1342222091  & 5h38m34,92s  -69d06m07,0s &\NII $_{122}$                     & 740                             & 452,0                                  &  1  &  faint line, pointed\\ % &1&  1     
     \hline  
     1342222092  & 5h38m45,00s  -69d05m23,0s &\OI $_{63}$                        & 740                             & 1734,0                                &  6  &  faint line\\           % &  1  &  1 
     \hline  
     1342222093  & 5h38m35,00s  -69d05m39,0s  &\OI $_{145}$                       & 740                             & 1321,0                               &  4  &  faint line\\    % &  1  &  1      
     \hline  
     1342222094  & 5h38m45,00s  -69d05m23,0s &\OIII $_{88}$,  \CII $_{158}$   & 740                             & 3325,0                                &  6  &  faint line\\      % &  1  &  1     
     \hline  
     1342222095  & 5h38m46,10s  -69d04m58,8s  &\NII $_{122}$                    & 740                             & 452,0                                 &  1  &  faint line, pointed\\ % &1 &  1       
     \hline  
     1342222096  & 5h38m45,00s  -69d05m23,0s  &\OI $_{145}$                      & 740                             & 1735,0                              &  6  &  faint line\\    % &  1  &  1        
     \hline  
     1342222097  & 5h38m35,00s  -69d05m39,0s  &\OI $_{63}$                        & 740                             & 1320,0                              &  4  &  faint line\\    % &  1  &  1      
     \hline  
     1342231279  & 5h38m38,00s  -69d06m00,0s  &\CII $_{158}$                     & 889                            & 801,0                                 &  3  &  bright line\\  % &  1  &  1       
     \hline  
     1342231280  & 5h38m30,00s  -69d06m07,0s  &\OIII $_{88}$                      & 889                             & 724,0                                &  3  &  bright line\\   % &  1  &  1     
     \hline  
     1342231281  & 5h38m30,00s  -69d05m07,0s  &\CII $_{158}$                     & 889                             & 801,0                                &  3  &  bright line\\    % &  1  &  1       
     \hline  
     1342231282  & 5h38m55,00s  -69d03m49,0s  &\OI $_{63}$ , \CII $_{158}$       & 889                             & 1133,0                              &  2  &  bright line\\  % &  1  &  1          
     \hline  
     1342231283  & 5h38m30,.00s  -69d06m07,0s  &\OI 63, \CII $_{158}$       & 889                             & 1420,0                             &  3  &  bright line\\     % &  1  &  1      
     \hline 
     1342231284  & 5h38m56,00s  -69d04m50,0s  &\NII $_{122}$                     & 889                             & 662,0                                &  2  &  bright line\\   % &  1  &  1     
     \hline  
     1342231285  & 5h38m40,00s  -69d04m38,0s  &\OIII $_{88}$                      & 889                            & 576,0                                 &  2  &  bright line\\   % &  1  &  1       
     \hline  

\end{tabular} 
\caption{Technical details on the observations. All of the observations were done in unchopped mode.}
\label{tab:observations}
\end{center}
\end{table}
}
\end{landscape}

\section{Spherical geometry}
\label{app:wrap}

\begin{figure*}
     \centering
      \includegraphics[trim= 7mm 3mm 5mm 5.5mm, clip, width=8cm]{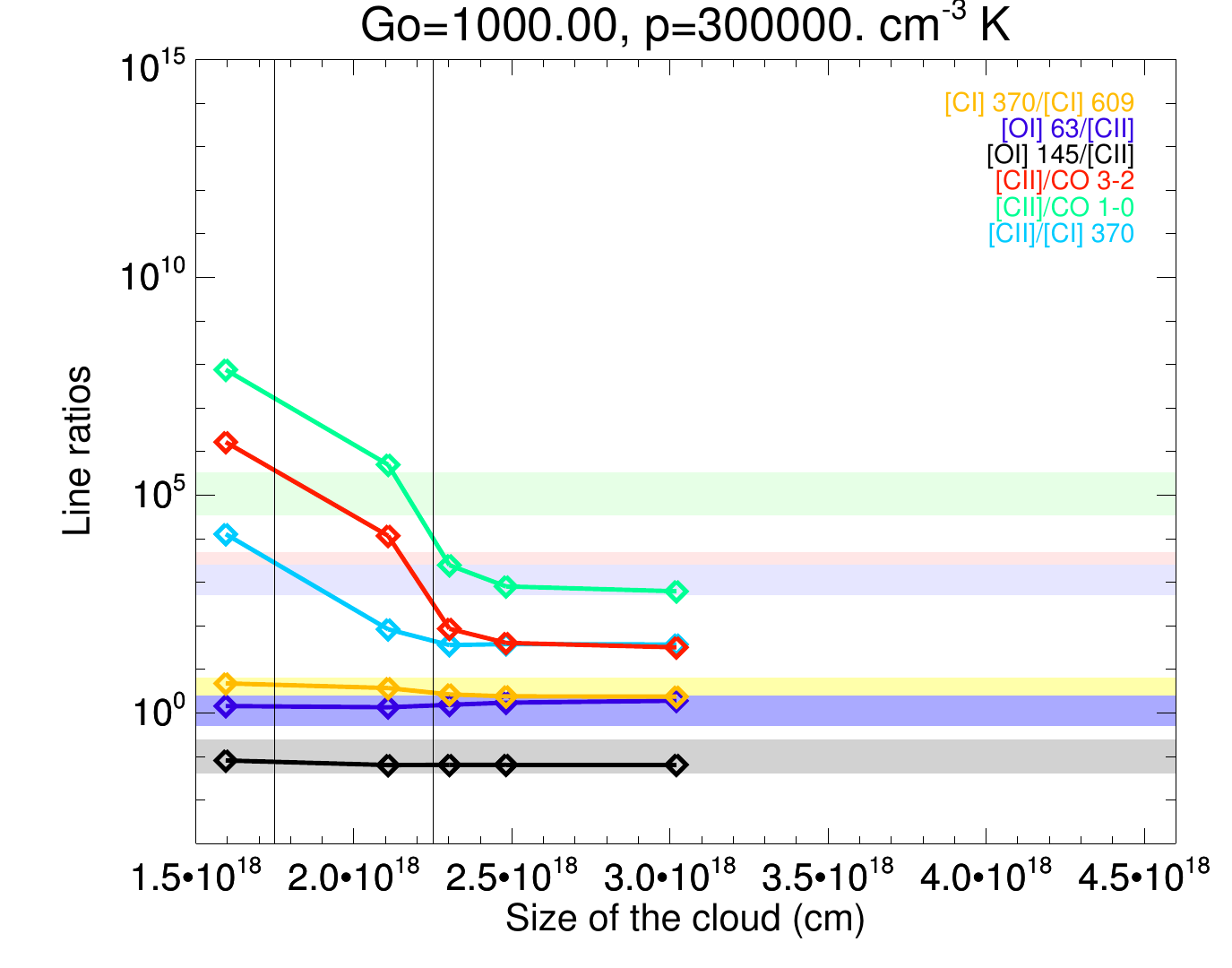}
      \hspace{0.5cm}
      \includegraphics[trim= 7mm 3mm 5mm 5.5mm, clip, width=8cm]{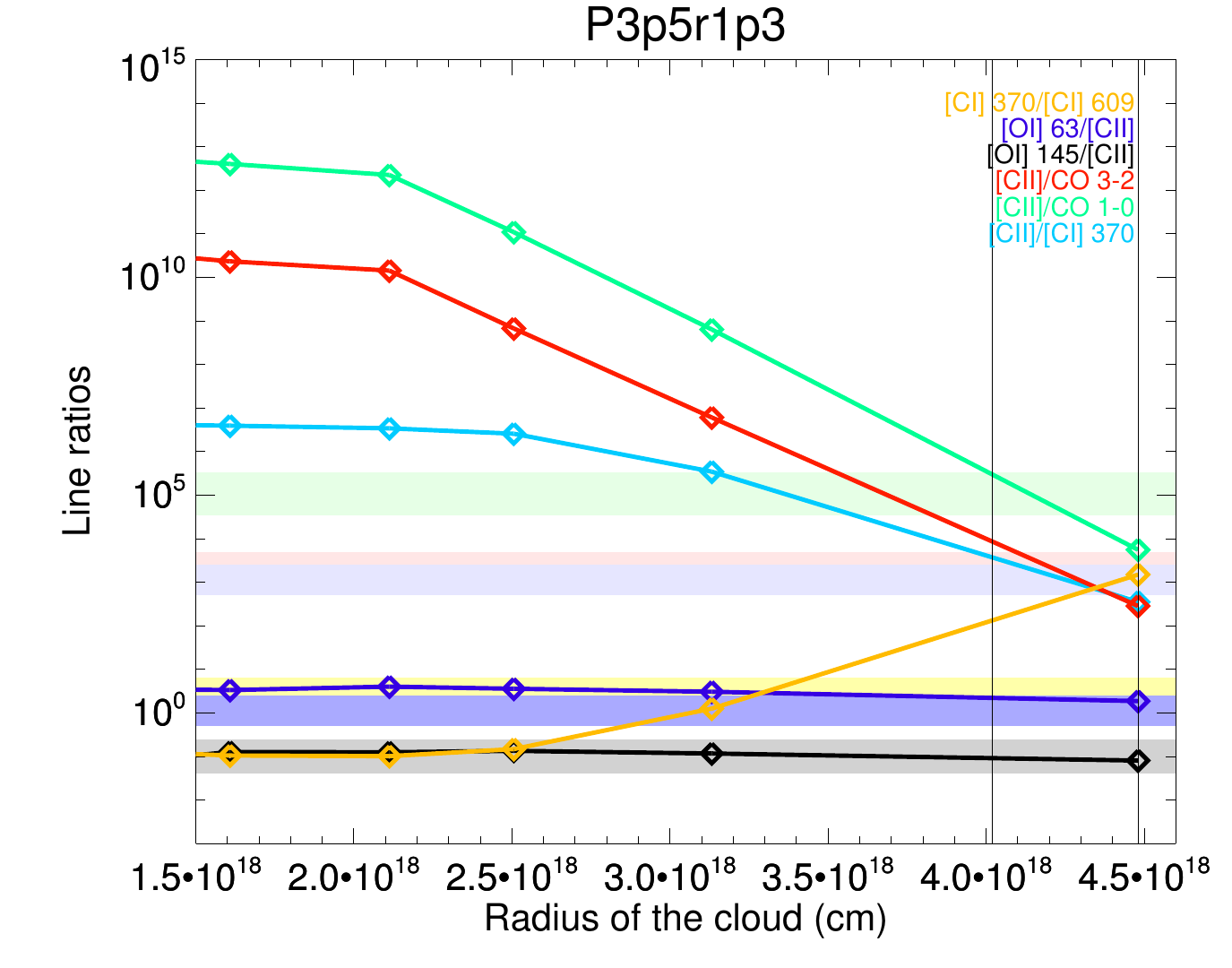}
       \caption{\textit{Left:} Ratios $\mathcal{R}$ of the modeled line ratios for \CI ~370\mic/\CI ~609\mic , \OI ~63\mic/\CII, \OI ~145\mic/\CII , \CII /CO(3-2), \CII / CO(1-0) and \CII / \CI 370\mic , for simulated plane parallel clouds of different sizes, with \go\ $=1\times10^3$ and P $=3\times10^5$ \cm ~K.
                                  \textit{Right:} Same for spherical clouds of different radius.
      The vertical lines indicate the range of radius where the predictions of the model are compatible with the observed \CII / CO(3-2) and \CII / \CI 370\mic . 
                       }
              \label{fig:wrap}
    \end{figure*}

The Meudon PDR code is a plane parallel model.
The code computes the abundance profiles of the various species and excited states in a plane parallel system as a function of the distance to the surface of the cloud.
It is possible to post-process the results of a simulation to wrap the structure and simulate a spherical cloud.
To do that, we integrate the intensity of each transition over a sphere, where the abundance profiles of each species as a function of the distance to the surface of the sphere is equal to the computed abundance profile as a function of the distance to the surface for a plane parallel geometry.
The resulting line ratios as a function of the diameter of the sphere, for an integrated cloud illuminated by an isotropic field, are shown in the right panel of Figure~\ref{fig:wrap}.

This is approach is geometrical. 
It is not done to accurately model the physics of a spherical cloud but to investigate the impact of the geometry on the integrated intensity, similar to the approach or \cite{Bolatto1999}.
Physically, this approach is only valid when there is enough extinction, i.e. for values of \Av\ $\gtrsim 5$.
In addition, we wish to emphasize that the approach of computing a spherical cloud by wrapping the structure is not satisfactory in the case of clouds illuminated by a central stellar cluster since the radiation field seen by any cloud is not isotropic. 
The ideal model would be a model with clouds of various sizes located at various distances from a central radiation source.

\bibliographystyle{aa} % style aa.bst 
\bibliography{PDR30Dor} % your references Yourfile.bib

\end{document}